\documentclass[10pt,journal,compsoc]{IEEEtran}
\ifCLASSOPTIONcompsoc
  \usepackage[nocompress]{cite}
\else
  \usepackage{cite}
\fi

\usepackage{graphicx}
\ifCLASSOPTIONcompsoc
  \usepackage[caption=false,font=footnotesize,labelfont=sf,textfont=sf]{subfig}
\else
  \usepackage[caption=false,font=footnotesize]{subfig}
\fi

\usepackage[english]{babel}
\usepackage{hyphenat}

\usepackage{url}

\usepackage{balance}

\usepackage{booktabs}
\usepackage{array}
\ifCLASSOPTIONcaptionsoff
  \usepackage[nomarkers]{endfloat}
 \let\MYoriglatexcaption\caption
 \renewcommand{\caption}[2][\relax]{\MYoriglatexcaption[#2]{#2}}
\fi

\usepackage[ruled,linesnumbered]{algorithm2e}
\usepackage{amsmath,amsfonts}

\usepackage{xcolor}

\usepackage[driverfallback=dvipdfm,bookmarksopen=true,colorlinks,citecolor=black,linkcolor=black,anchorcolor=black,urlcolor=black]{hyperref}

\usepackage{fixltx2e}

\begin{document}

\title{ClassyTune: A Performance Auto-Tuner for Systems in the Cloud}

\author{\IEEEauthorblockN{Yuqing~Zhu,~\IEEEmembership{Member,~IEEE}, and Jianxun~Liu\\}

\IEEEcompsocitemizethanks{\IEEEcompsocthanksitem Y. Zhu is the corresponding author. E-mail:zhuyuqing@ict.ac.cn.
\IEEEcompsocthanksitem Y. Zhu is with the Institute of Computing Technology, Chinese Academy of Sciences, Beijing 100190, China.
\IEEEcompsocthanksitem J. Liu is with UTuned Technology Company Limited, Beijing, China.}
\thanks{Manuscript received 25 Dec. 2018; revised 18 Jun. 2019.}}

\markboth{IEEE Transactions on Cloud Computing, June~2019}%
{Zhu \MakeLowercase{\textit{et al.}}: ClassyTune: A Performance Auto-Tuner for Systems in the Cloud}
%

\IEEEtitleabstractindextext{%
\begin{abstract}
Performance tuning can improve the system performance and thus enable the reduction of cloud computing resources needed to support an application. Due to the ever increasing number of parameters and complexity of systems, there is a necessity to automate performance tuning for the complicated systems in the cloud. The state-of-the-art tuning methods are adopting either the experience-driven tuning approach or the data-driven one. Data-driven tuning is attracting increasing attentions, as it has wider applicability. But existing data-driven methods cannot fully address the challenges of sample scarcity and high dimensionality simultaneously. We present ClassyTune, a data-driven automatic configuration tuning tool for cloud systems. ClassyTune exploits the machine learning model of \emph{classi}fication for auto-tuning. This exploitation enables the induction of \emph{more} training samples \emph{without} increasing the input dimension. Experiments on seven popular systems in the cloud show that ClassyTune can effectively tune system performance to seven times higher for high-dimensional configuration space, outperforming expert tuning and the state-of-the-art auto-tuning solutions. We also describe a use case in which performance tuning enables the reduction of 33\% computing resources needed to run an online stateless service.\vspace{-6pt}
\end{abstract}

}

\maketitle

\IEEEdisplaynontitleabstractindextext

%
\IEEEpeerreviewmaketitle


\IEEEraisesectionheading{\section{Introduction}\label{sec:introduction}}

\begin{figure}[t]
\vspace{-12pt}
 \subfloat[MySQL performance under different workloads.]{
       \includegraphics[width=.22\textwidth,height=71pt]{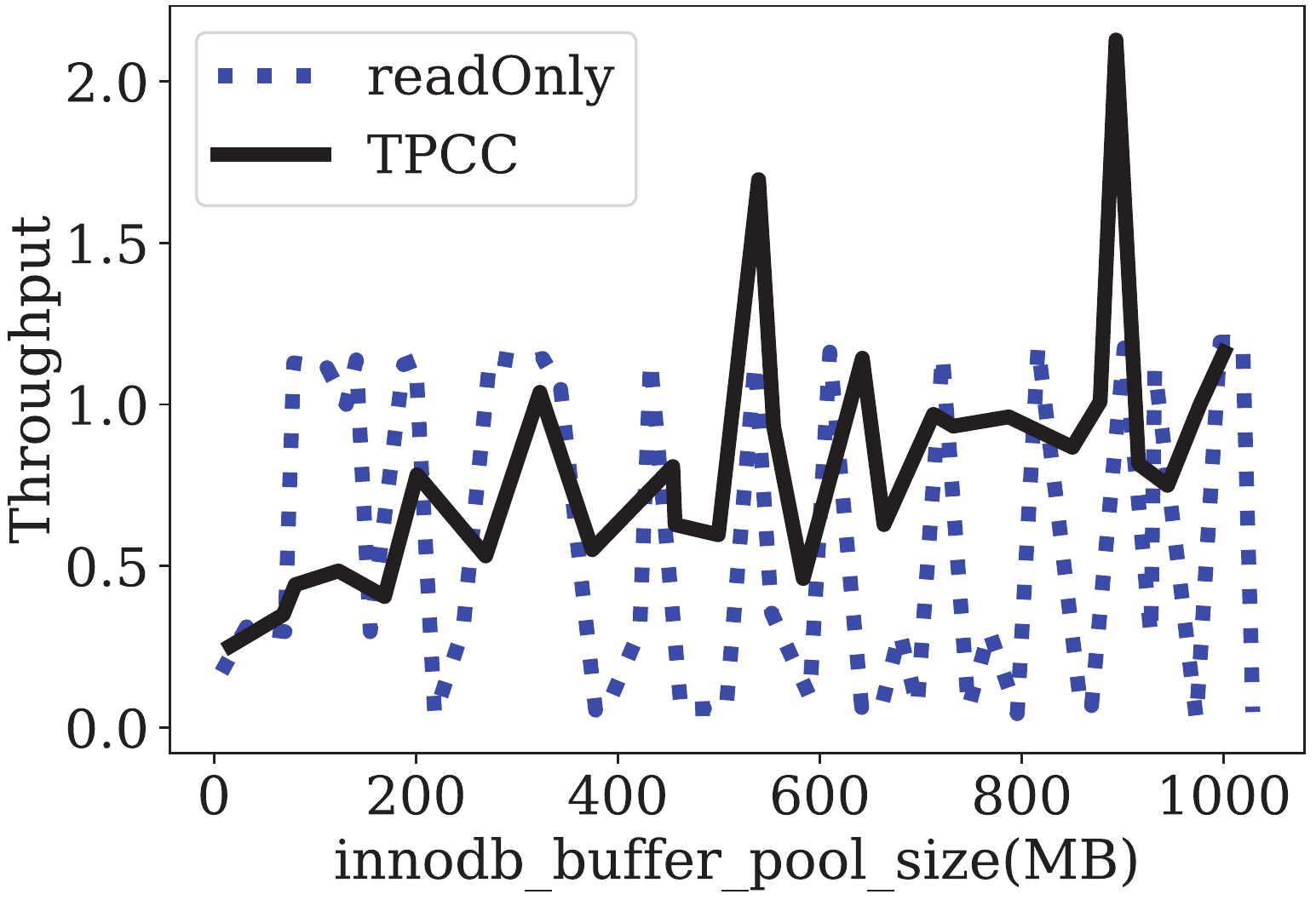}%
        \label{fig:curve:mysql} 
    }\vspace{-3pt}
\hspace{1pt}
  \subfloat[Spark performance under different environments.]{
        \includegraphics[width=.22\textwidth,height=71pt]{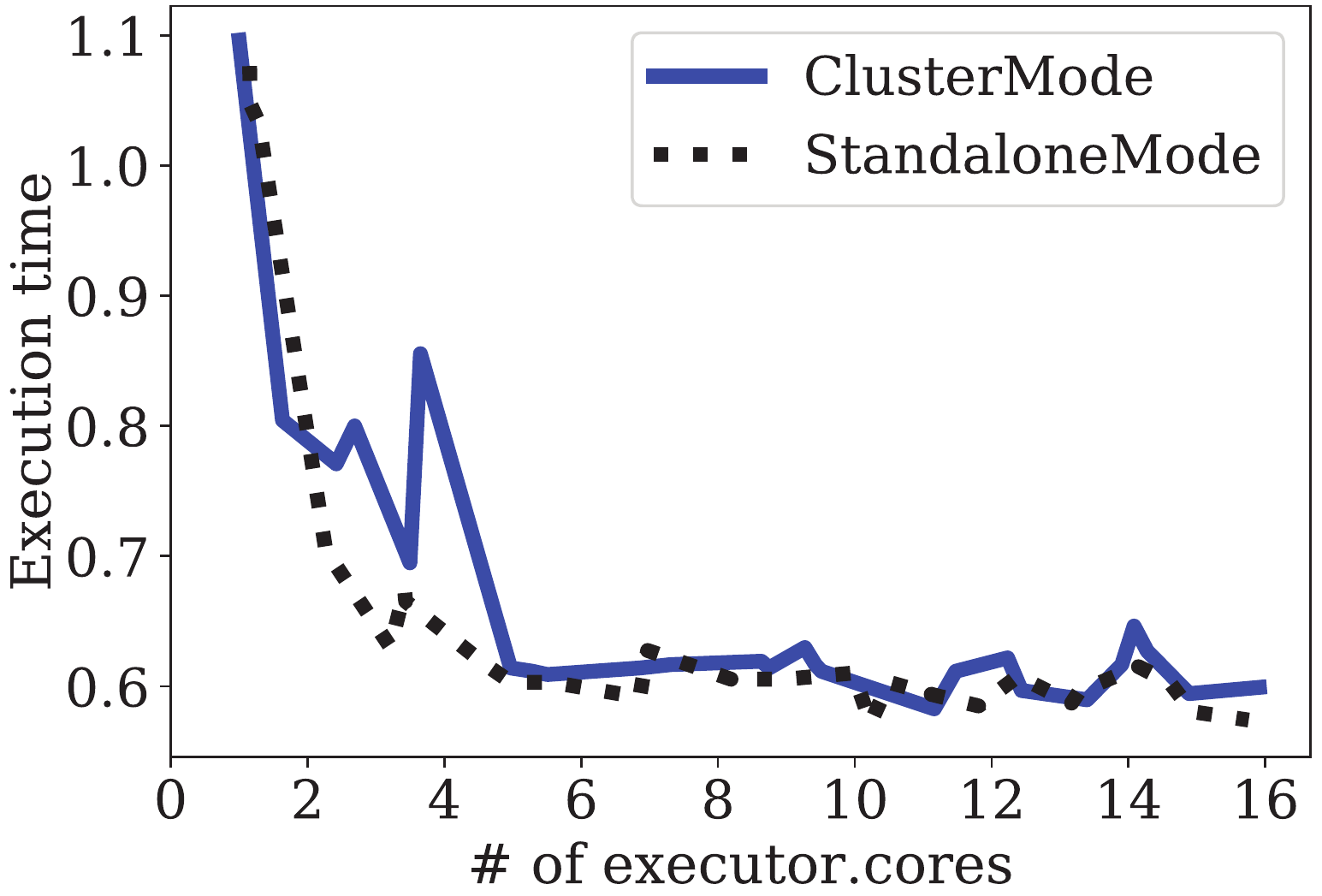}%
        \label{fig:curve:spark} 
    }\vspace{-3pt}
     \caption{Performance-PerfConf curves are \emph{nonlinear}, \emph{nonsmooth}, and \emph{system-/workload-/environment-specific}.}\vspace{-6pt}
     \label{fig:curve}
\end{figure}

\IEEEPARstart{C}{loud} computing has facilitated the deployment of systems for big data analytics and Web services. For an efficient exploitation of the cloud computing resources, we can either choose for a specific task~\cite{tccPerf} the most cost-effective cloud configuration, i.e., the types and numbers of virtual machine instances; or, we can optimize the system performance for a specific deployment setting so as to reduce the total computing resources in demand~\cite{bestconfig}. In fact, modern systems are exposing an increasing number of configurable parameters that can have strong impacts on system performance and thus that are denoted as \emph{PerfConfs}, e.g., \emph{innodb\_buffer\_pool\_size} and \emph{executor.cores} in Figure~\ref{fig:curve}. Well tuning the PerfConfs of a system can lead to multiple times of performance speedup~\cite{acts}, requiring no change to the system design. Unfortunately, to meet the diversity of applications and deployment settings, the number and the complexity of PerfConfs have increased to a level exceeding the comprehension capability of human beings~\cite{asilomar}. We see an emerging need for automating the tuning of PerfConfs~\cite{ieee1,metis,boat} for much higher system performance.

Existing solutions to auto-tuning PerfConfs for systems in the cloud are either \emph{experience-driven} or \emph{data-driven}. Approaches based on heuristics-guided search~\cite{ga,evoParatune} and analytical modeling~\cite{starfish,tccEnsemble} rely heavily on human experiences and knowledge, belonging to the experience-driven category. Experience-driven tuning requires human intervention for each specific case and has limited applicability. Approaches using Bayesian optimization~\cite{ottertune} or other machine-learning models~\cite{capes} exploit data to train models for optimization, thus falling into the data-driven category. Data-driven tuning can be applied to where sufficient tuning samples are provided, thus attracting increasing popularity~\cite{boat,opentuner}. However, running tuning tests in the cloud and collecting large samples are expensive due to a pay-as-you-go cost, while the sample size required is in proportion to the dimension of the configuration space~\cite{learningTheory}. Sample scarcity and high dimensionality place two challenges to data-driven configuration tuning.

In this paper, we take the data-driven approach to address the problem of auto-tuning performance for systems in the cloud through adjusting PerfConf settings. Our main idea is to \emph{tackle performance tuning as a comparison problem and model the performance comparison relations of the limited samples}. In contrast to the common exploitation of performance prediction models~\cite{cartModel}, we adopt the classification method for the comparison modeling, as it can bring about two benefits that directly address the sample scarcity challenge of the data-driven tuning. First, the classification model for the comparison problem can have a training set as quadratically large as the original sample set, as it takes pairs of original samples as input and such pairs can be constructed through \emph{permuting} every pair of the original samples. Second, we can generate even more training samples based on manual tuning experiences. As manual tuning process usually goes through numerous trials and comparisons, the tuning experiences are usually summarized in comparison-based rules, e.g., \textit{increasing memory cache sizes leads to higher performances}. We can generate more training samples for the classification models based on such tuning rules, while this is impossible for the performance prediction modeling~\cite{rf}.

But two problems remain to be solved. The first is about dimensionality, i.e., \emph{how to effectively represent the input without increasing its dimensions}. If we directly concatenate two PerfConf settings, the input dimension for the model is increased to twice of the original one, leading again to sample scarcity~\cite{learningTheory}. If we take the division or difference of a PerfConf pair, different pairs will collide, leading to different inputs mapped to one same output. We propose to induce samples by constructing a bijection from a $2d$-dimensional space to a $d$-dimensional one. The second is about model accuracy, i.e., \emph{how to find the best PerfConf setting using an imprecise model}. This found PerfConf setting should lead to the best system performance within a given time and computing resources. Machine learning models are generally not a hundred percent accurate~\cite{learningTheory}. Even if we train a model with enough samples, this model can still mistakenly distinguish some comparison relations. We must robustly find a best PerfConf setting even if some predictions are incorrect. This best PerfConf setting should lead to better performance. We propose a clustering-based tuning algorithm that can exploit the imprecise classification model.

%

We thus present ClassyTune, which is, to the best of our knowledge, the first automatic performance tuning system that exploits a classification model to find the best PerfConf setting within a limited sample. In Classytune, we use a \emph{classifier} to predict whether one PerfConf setting has a better performance than another. Taking this classification approach, ClassyTune can construct a useful model for auto-tuning with only a limited number of original PerfConf-performance samples, while the common auto-tuning methods would require tens of times more samples~\cite{smartconf,ottertune,bestconfig}. The classifier model can make a prediction in a time multiple orders of magnitude shorter than a tuning test actually runs. We can thus use the model as the surrogate of the system and take a systematic approach towards tuning with an imprecise model.

ClassyTune consists of three components for sampling, modeling and searching respectively. The sampling component outputs a database of PerfConf-performance samples; the modeling component outputs a classification-based model; and, the searching component finds the PerfConf setting with the highest performance in best effort. Decoupling the system into three components allows the reuse of the intermediate tuning outputs, i.e., the database and the model. As a result, ClassyTune can be used not only for tuning, but also for system analysis. The intermediate outputs, especially the model, can inform users about relations between PerfConfs and performance,

In this work, we make the following contributions:\vspace{-3pt}
\begin{itemize}
  \item We propose a data-driven performance auto-tuning approach, unprecedentedly adopting a classification model for representing the performance comparison relations between PerfConf settings ($\S$\ref{sec:modeling}).\vspace{3pt}
 \item We propose to address the input dimension problem through sample induction that constructs a bijection based on the Cantor's proof ($\S$\ref{sec:sampleMapping}).\vspace{3pt}
 \item We propose a clustering-based auto-tuning method that exploits the imprecise classification model ($\S$\ref{sec:tuning}).\vspace{3pt}
 \item We implement the above solutions in ClassyTune ($\S$\ref{sec:implement}) and evaluate the system in extensive and comprehensive experiments, using 7 popular systems and 14 common application workloads in the cloud ($\S$\ref{sec:eval}).\vspace{3pt}
 \item We present a customer's use case to show how ClassyTune can be used and help users reduce the cloud computing resources needed to run an online stateless service ($\S$\ref{sec:usecase}).
\end{itemize}

\section{Motivation and Related Work}%
\label{sec:motivate}%

This section examines the modeling challenges for the data-driven methods of automatic performance tuning based on PerfConf setting adjustments. These challenges motivate our work over the related works, which are summarized at the end of this section.

\subsection{Challenge: Non-Smooth Complicated Curves}%
\label{sec:curve}%

PerfConf-performance curves are formed by taking PerfConfs as input and the system performance as output. Different systems have different performance curves. In fact, this curve is not only related to the system, but also very sensitive to the workloads, the deployment environments and the computing resources~\cite{acts}. Figure~\ref{fig:curve} plots the curves for database system MySQL and the distributed online processing system Spark.

Among the four plotted curves, two for MySQL and two for Spark, none demonstrates linearity. The performance is not in direct proportion to the PerfConf input. For example, Figure~\ref{fig:curve:mysql} plots the throughput of MySQL under two workloads of read-only and TPC-C, given \texttt{\small buffer\_pool\_size} as input. The throughputs of MySQL are not directly proportional to the size of buffer pool. Figure~\ref{fig:curve:spark} plots the job durations for Spark under the standalone and  cluster deployments respectively. The performances demonstrate no linearity with the number of executor cores either.

Even for the same system, changes to the workload, the deployment environment, or the computing resources can also lead to different PerfConf-performance curves. Changing the workload from read-only to TPC-C leads to two completely different performance curves for MySQL, as shown in Figure~\ref{fig:curve:mysql}. Changing the deployment from the standalone mode to the cluster mode also changes the shape of Spark's performance curve, as illustrated in Figure~\ref{fig:curve:spark}.

Generally, it would not be wise to use linear models to map PerfConf-performance relations due to non-linearity. As the system, workload, environment, and computing resources are factors influencing the curve shape, PerfConf-performance models should be constructed with regard to a specific \emph{combination} of these factors, making model reuse infeasible. In sum, tuning tests and samples must be collected specifically for such a combination, leading to the sample scarcity challenge ($\S$\ref{sec:scarcity}).

The non-smooth property of performance curves is also a challenge to the popular data-driven auto-tuning method of Bayesian optimization (BO)~\cite{BO}. BO-based auto-tuning enables an effective use of the sampling budget by guiding the sampling and search process with an acquisition function. The common application of BO adopts a Gaussian process prior to get a closed-form acquisition function. Unfortunately, this adoption requires the objective function to be a differentiable function. But not all objective function is differentiable. In fact, it is shown that the performance surfaces of several popular cloud systems are non-differentiable~\cite{acts}. The dissatisfaction of this assumption can invalidate an optimization process based on BO.

\subsection{Challenge: The Sample Size}
\label{sec:scarcity}

Data-driven auto-tuning methods commonly exploit machine learning algorithms for modeling. We illustrate the sample size challenge to the common performance-prediction based modeling~\cite{cartModel}. We model the PerfConf-performance relation by three machine learning methods. As performance is a continuous value, these models are \emph{regression} models, including boosted decision tree (B\_CART), supported vector regression (SVR) and random forest regression (RFR). The decision tree model CART is effective in performance modeling for simple systems~\cite{cartModel} and thus recently applied to performance tuning~\cite{perfRank}. SVR can increase the sample set to twice as large, alleviating the sample scarcity partially. As a robust ensemble model, RFR combines the advantages of statistical reasoning and machine learning approaches~\cite{rf}. We have also tried linear regression, which has been used in a state-of-the-art related work for feature selection~\cite{ottertune}, but the model is too imprecise to be useful due to the reason described in Section~\ref{sec:curve}.

We measure the above models using the max prediction error, which is the max difference between the real performances and the model predictions, divided by the corresponding real performance. The equation is $\max(\{\frac{|y_i-\hat{y}_i|}{y_i}\}_{i\in[0,n-1]})$, where $n$ is the number of samples, $y$ is the real performances and $\hat{y}$ is the performances predicted by a model. We use 100 samples to construct each model over 10 PerfConfs.

\begin{figure}[t]
\hspace{-1pt}
 \subfloat[Max differences between predicted and real performances.]{
       \includegraphics[width=.255\textwidth,height=73pt]{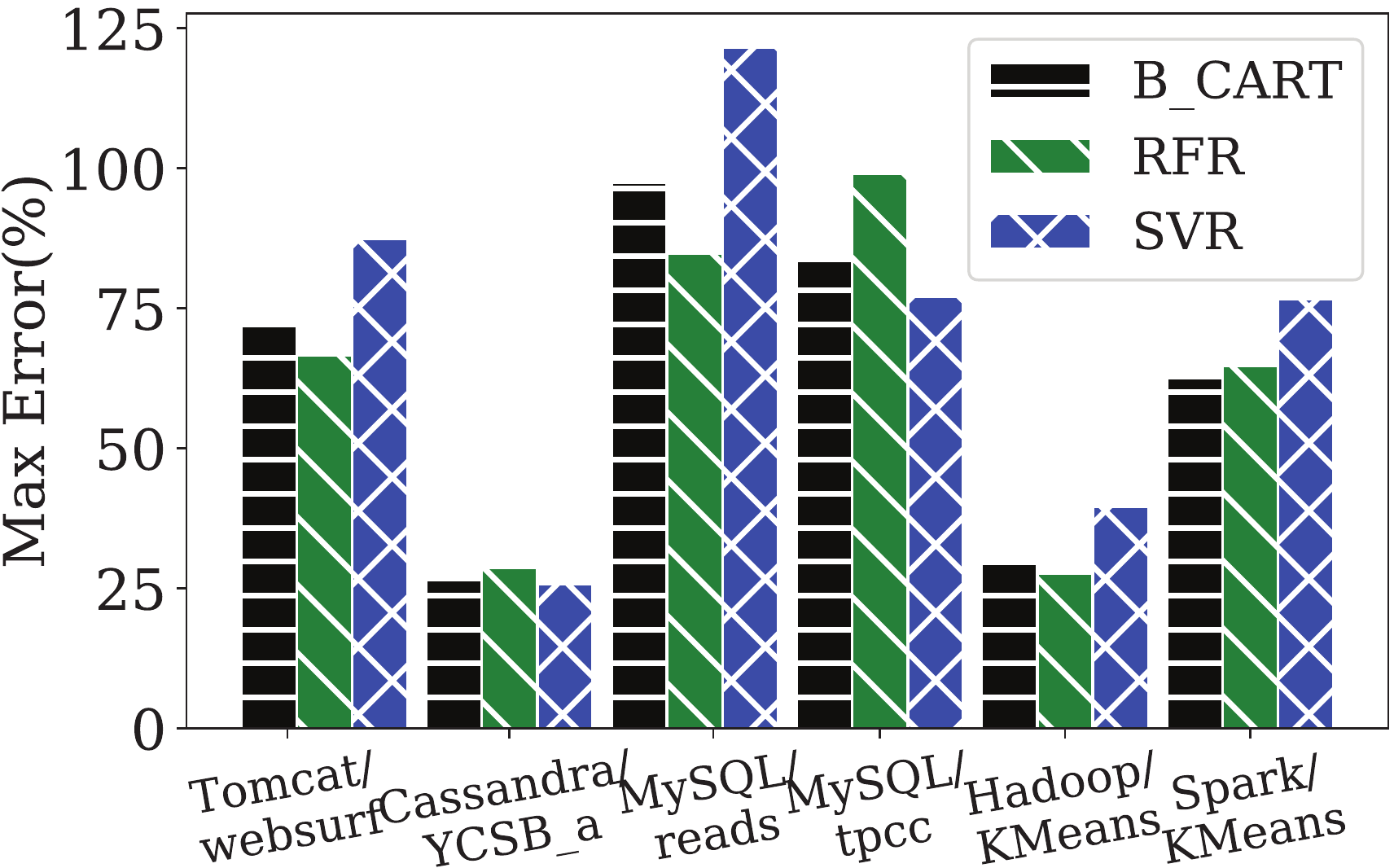}%
        \label{fig:sample:maxdiff} 
    }\vspace{-3pt}
\hspace{1pt}
  \subfloat[Errors reduced as samples added (Hadoop-KMeans/RFR).]{
        \includegraphics[width=.195\textwidth,height=73pt]{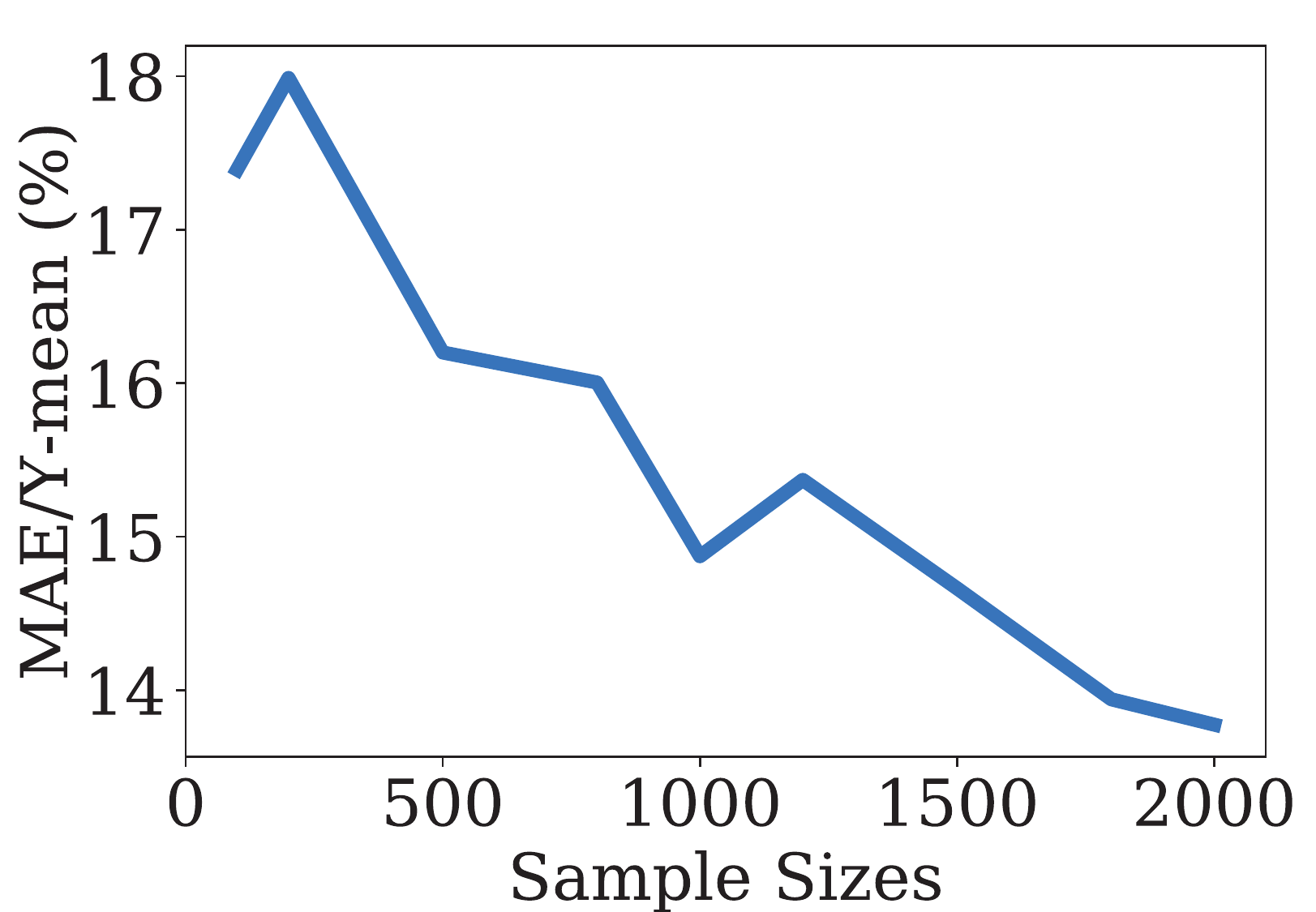}%
        \label{fig:sample:sizes} 
    }\vspace{-3pt}
     \caption{Highly inaccurate performance predictions \textbf{due to limited samples}, but adding samples reduces errors.}\vspace{-12pt}
     \label{fig:sample}
\end{figure}
As demonstrated in Figure~\ref{fig:sample:maxdiff}, the max prediction errors of these models can be very high, as much as twice more than the real performances. While the complexity of the PerfConf-performance curves is one reason, the scarcity of samples is the other. In fact, the model inaccuracy can be decreased given more samples (Figure~\ref{fig:sample:sizes}), but the cost of obtaining a large sample set can be high. Many tuning solutions require a database of thousands of samples for tuning 10 parameters~\cite{ottertune}. Models based on neural networks would require more samples even for just two PerfConfs~\cite{capes}. Worse still, these samples must be collected for each specific combination of system, workload, environment and computing resources. This makes the precise prediction on system performance almost impossible, because collecting a large number of samples for \emph{every} such combination is impractical, if not impossible. Hence, we are facing the problem how to obtain proper samples for model construction.

\begin{figure}
\hspace{-1pt}
 \subfloat[Better optimized performances for a larger initial training set.]{
       \includegraphics[width=.22\textwidth,height=73pt]{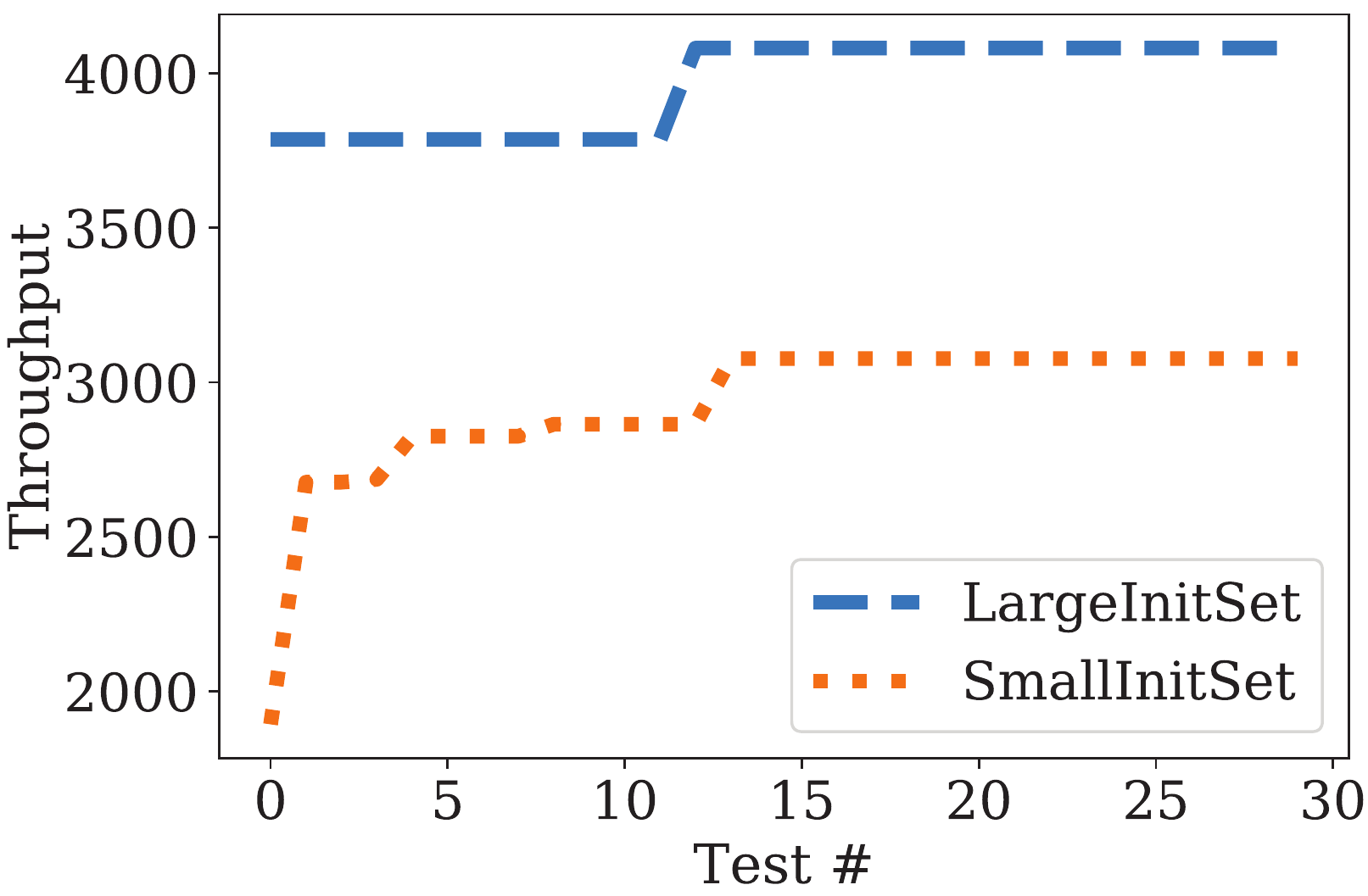}%
        \label{fig:gpr:best} 
    }\vspace{-3pt}
\hspace{1pt}
  \subfloat[Better next point predictions for a larger initial training set.]{
        \includegraphics[width=.22\textwidth,height=73pt]{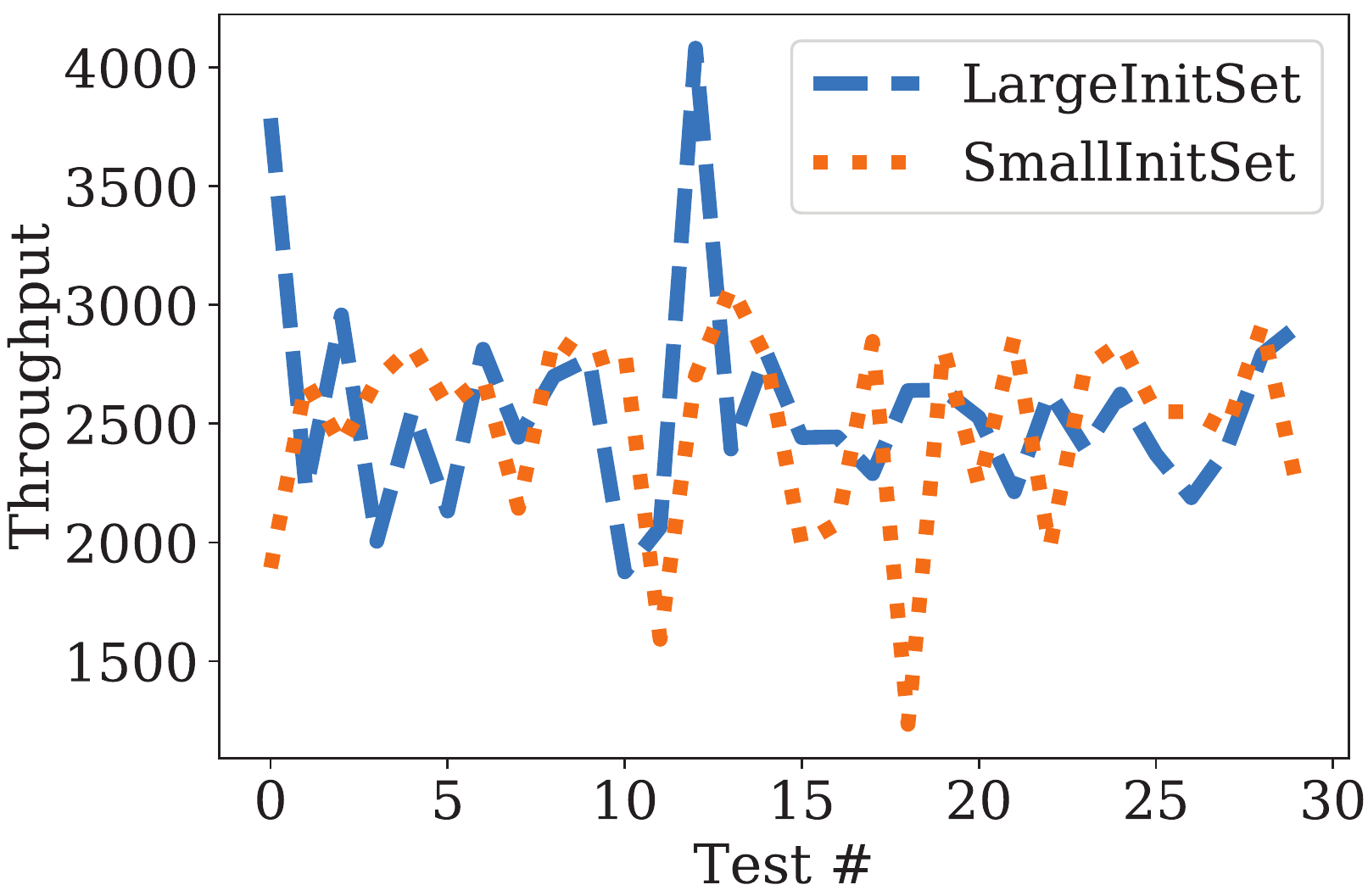}%
        \label{fig:gpr:next} 
    }\vspace{-3pt}
     \caption{\textbf{Sample size matters:} tuning Tomcat by BO with a GP prior.}\vspace{-6pt}
     \label{fig:gpr}
\end{figure}

Sample scarcity has also negative impacts on the tuning process of BO. With BO, the GP model can be trained with limited samples and later updated with more samples as the acquisition function drives the sampling process. However, with a GP model trained with limited samples, the tuning process based on BO can be very ineffective. As demonstrated in Figure~\ref{fig:gpr}, a BO model with very few samples cannot locate best points for sampling as one with more samples does.
\subsection{Challenge: Irrelevant Optimization Steps}

Data-driven tuning methods like Bayesian optimization optimize and sample stepwise towards the final optimization goal~\cite{ottertune}. In comparison, many other data-driven tuning methods train a model after taking a large samples and then optimize on the final model~\cite{perfRank}. There exists a question on\emph{ whether we should optimize stepwise or integrally}.

We look into the optimization process of BO. At each step, BO algorithms determine the next sampling point by \emph{optimizing} a carefully designed acquisition function~\cite{BO}. Acquisition functions determine how to explore the input space. The commonly used acquisition function is the expected improvement (EI) function, which represents the expected improvement on sampling a given point. The prior probability model on $f$ is needed in the EI computation. This probability model is usually assumed to be described by a Gaussian process (GP)~\cite{ottertune}. Assuming the GP prior, a priori knowledge over $f$ is required to set the covariance function and hyper-parameters. We take the common practice in the choice of the covariance function and hyper-parameters~\cite{BO}.

Figure~\ref{fig:gpr:next} demonstrates how the BO method runs toward the final result by optimizing the EI acquisition function at each step. Even though the current EI acquisition function is optimized to find the next sample point at every step, the found point is not necessarily a better one. In fact, it is a worse one in many cases as demonstrated in Figure~\ref{fig:gpr:next}. When the total number of samples is small, the resulting model might even fail to find a better point in the following steps, e.g., the optimization process with a small initial sample set as represented by the dotted line in Figure~\ref{fig:gpr}.

These facts indicate that we do not need to optimize at every step in the optimization process. We can wait till enough samples are collected. We should optimize integrally on a large sample set, instead of on a small sample set and in a stepwise way. Besides, instead of trying on a single point at each step, we can simultaneously try multiple points. With these understandings, we design ClassyTune.

\subsection{Related Work}

Solutions to automatic performance tuning have been proposed for a specific type of systems, e.g., storage systems~\cite{capes} and databases~\cite{ottertune}. Auto-tuning for general systems also exist, e.g., BestConfig~\cite{bestconfig}, BOAT~\cite{boat}, and SmartConf~\cite{smartconf}. Performance tuning requires the support from a flexible system architecture. Thus, auto-tuning systems for general systems implement system architectures for supporting the whole process of auto-tuning PerfConfs, including manipulating the system under tune, running tuning tests and computing the optimization results. At the core of configuration tuning lies a black-box optimization problem. The solutions to this black-box optimization problem can be divided into two categories, i.e., \emph{experience-driven tuning} and \emph{data-driven tuning}.

Classic experience-driven tuning methods include the heuristics-based search approach~\cite{bestconfig,ga,sa} and the control-theory based approach~\cite{smartconf}. The tuning based on manually specified models~\cite{starfish,tccEnsemble} also belongs to this category. While heuristics are highly related to human experience, they might be useful for some systems but not the others. Besides, the search-based approach can only produce stable results when the searched space is large enough. Control theory based auto-tuning iteratively applies a change to inputs and monitors feedbacks to decide for the next step. This approach is only applicable to cases where the number of PerfConfs is only a handful. There also exist auto-tuning tools that decide the configuration settings based on expert provided guidelines or experts' answers on a set of questions~\cite{questiontune1}. Like manually specified models, they have only limited applicability. Different heuristics-driven tuning methods can be assembled for usage in auto-tuning, as the OpenTuner framework does~\cite{opentuner}.

Data-driven tuning approaches exploit data to guide tuning, instead of experience-based heuristics or manually specified models. Such approaches typically train a model on a given data set and optimize the model towards the tuning objective~\cite{ottertune}. Due to the large number of PerfConfs, the model-based approach demands a large sample set to train useful regression models on performance~\cite{learningTheory}. Bayesian optimization is a popular data-driven tuning approach~\cite{capes,ottertune,metis}, as it requires only a limited number of samples to train the optimization model. For the BO method with a GP prior, a priori knowledge over the black-box function is required to set the covariance function and hyper-parameters of the GP model. Unfortunately, such knowledge requires deep understanding of the optimization problem and the covariance function, which is a difficult task for common users. Facebook's Spiral system~\cite{spiral} is an industrial practice to integrate data-driven methods for predicting the current best setting of PerfConfs. A recent work BOAT~\cite{boat} enables the blending of experience-driven tuning and data-driven tuning. It proposes an optimization framework to integrate human knowledge into the Bayesian optimization process, making the black-box optimization partially white.

ClassyTune takes a classification approach to performance auto-tuning, which is completely different from previous works. ClassyTune addresses the sample scarcity problem in auto-tuning by two measures, i.e., permuting sample pairs to form inputs and generating samples from tuning experiences. Through data generation, ClassyTune transfers expert knowledge and experiences to the auto-tuning process. Like BestConfig~\cite{bestconfig}, ClassyTune has an architecture that can work with both experience-driven and data-driven tuning methods. The difference of these two architectures is that ClassyTune can save all collected tuning samples for future modeling purpose and expose the tuning model to inform users about PerfConf-performance relations, while BestConfig cannot. The classification model can be used effectively as the surrogate of the system in analysis. In comparison, models directly predicting performances are too imprecise to rely upon~\cite{perfRank}, while models like Bayesian optimization~\cite{BO,metis} can only predict the next best points and not be used in such analysis.

\section{Design Overview}

ClassyTune is a data-driven performance auto-tuning tool for systems in the cloud. It addresses the problem of auto-tuning system PerfConfs within a given number of tuning tests. A set of PerfConf-performance samples can be collected from the given number of tuning tests.

Taking a comparison-based perspective, Classytune models the relation between each pair of PerfConf-performance samples. This comparison-based modeling enables the generation of even more samples based on tuning experiences, further attacking the sample scarcity challenge. The modeling process trains a classifier for predicting whether the first PerConf setting has a higher performance than the second in a pair of PerfConf settings. Section~\ref{sec:modeling} presents the details of the comparison-based modeling based on classification. Unlike the performance-prediction based methods, ClassyTune does not need to assume whether the performance curve is linear or non-linear, thanks to its classification-based method. But, like other machine learning models, the trained classifier is not a hundred percent accurate. It is an imprecise classifier.

To tune with the imprecise classifier, ClassyTunes adopts a clustering-based method. Naive exploitations of the imprecise classifier will fail to find a best PerfConf setting due to occasionally incorrect predictions. ClassyTune uses the trained classifier as the surrogate of the system. ClassyTune clusters a set of good PerfConf settings output by the classifier to locate promising spaces for searching the best PerfConf setting. Section~\ref{sec:tuning} presents the details of the tuning process based on an imprecise classifier.

The overall architecture and implementation of ClassyTune is presented in Section~\ref{sec:implement}. ClassyTune consists of three main components, i.e., sampling, modeling and searching ($\S$\ref{sec:components}). Sampling and searching components can interact with the system under tune. The interaction is mainly adjusting the PerfConf settings of the system under tune. The interactions are automated and driven by the tuning process when needed. The three components interact in a way as defined by the tuning algorithm ($\S$\ref{sec:algo}).  Implementation details are also provided ($\S$\ref{sec:sysdetails}).\vspace{-6pt}

\section{Modeling Comparisons}
\label{sec:modeling}

In this section, we first formulate the comparison-based view for performance tuning. We then detail how to induce training samples and model comparison relations by classification for the auto-tuning task.\vspace{-6pt}

\subsection{The Comparison-Based View}

We model the performance-comparison relations between pairs of PerfConf settings. This comparison-based model takes a pair of PerfConf settings $(X_1, X_2)$ as input and outputs $1$ if the first setting has a performance better than the second, i.e., $f(X_1)-f(X_2)>0$, or $0$ otherwise. Hence, it can be represented by the function $g$ defined as:\vspace{-3pt}
\begin{equation}
g(X_1, X_2)=
\begin{cases}
1 & \text{if $f(X_1)-f(X_2)>0$},\\
0 & \text{otherwise}.
\end{cases}
\label{eq:classifier}
\end{equation}\vspace{-6pt}

We exploit the above comparison-based model to tackle the auto-tuning problem. We relate the comparison relation to \textbf{each dimension difference between an input pair}. We propose a mapping to encode this dimension difference and construct a new set of samples ($\S$\ref{sec:sampleMapping}). With the constructed sample set, we can exploit classifiers to model the comparison relations between input pairs. We choose the classifier with the best trade-off between computation overhead and accuracy for modeling.

This comparison-based approach has the following advantages over other approaches. First, the tuning problem can easily fit into the comparison-based modeling perspective. As performance tuning is usually formalized as an optimization problem, the problem for finding the optimal PerfConf setting is in fact to find one setting that has a better performance as \emph{compared} to all other PerfConf settings.

Second, modeling the comparison relations is more robust than directly modeling on performance. On sample collection, the performance measurements are in fact prone to noise, leading to a variance of measurements. But even if two measurements might not be accurate due to noise or fluctuation, their comparison result can still be correct. In case that some comparisons do not have correct results due to a high variance of measurements, there still exist many other correct comparison relations to rely upon. In comparison, such high variance of measurements can completely divert the modeling of performance predictions.

Third, comparison-based modeling leads to a natural augmentation of the data set, partially alleviating the sample scarcity problem. With comparison-based modeling, the training set consists of PerfConf pairs and their performance comparison results. This training set must be mapped from the original set of PerfConf-performance  samples. The mapping is a permutation of the original sample set. Thus, for the same sample collection effort, comparison-based modeling can have a training set as quadratically large as the direct modeling of performance can have. Besides, we can generate even more training samples based on manual tuning experiences, which are commonly expressed as comparison-based rules. This is impossible for the performance prediction modeling.

Finally,  the comparison-based modeling provides straight-forward means for users to gauge the influences of PerfConfs on the performance. On manual tuning, we would actually observe whether a change of PerfConf values leads to an increase or decrease of the performance. This is exactly a comparison process. In fact, when we make an analysis on systems, we make similar comparison-based observations as well. Thus, comparison-based modeling aligns well with the thinking of human beings.\vspace{-6pt}

\subsection{Inducing Samples for Modeling}%
\label{sec:sampleMapping}

The performance comparison result can be viewed as the performance change result if the first PerfConf setting is changed to the second one. Put it in another way, the performance change is actually related to the first PerfConf setting and the value difference regarding the second PerfConf setting. Hence, we can represent a pair of PerfConf settings by encoding in each dimension the value of the first setting and the corresponding difference respectively. For each dimension, we need to construct a bijection for an effective encoding. With such bijection, we can construct a larger sample set without increasing the input dimension.

Cantor's proof is the solution to constructing such bijection~\cite{sfcbook}. Probably sounding counter-intuitive, it has been shown in cardinal arithmetic that the cardinality of the set $[0,1]\times[0,1]$ (the unit square) is equal to that of the set $[0,1]$. The cardinality of a set is a measure the number of elements of the set. The cardinality of a set is also called its size. The cardinality of a finite set is the number of its elements. \emph{Two sets have the same cardinality if there exists a bijection between the two sets.} This result was first demonstrated by Cantor and later proved based on space-filling curves (SFC), which are curved lines twisting and turning enough to fill the whole of any finite space~\cite{sfcbook}. Space-filling curves provide one way for constructing a bijection from the unit square to the unit interval, mapping from the $2d$-dimension space to the $d$-dimension space.

For each PerfConf, we thus construct the bijection from two values into one value using SFC, specifically the z-ordering method~\cite{sfcbook}. The mapped value in the unit interval is called the z-value. The z-value of a point in multi-dimensions is simply calculated by interleaving the binary representations of its coordinate values. For example, given the $i$th-dimension values $X^{(i)}_1=0b000100$ and $X^{(i)}_2=0b000101$, we can get the z-value of $(X^{(i)}_1,X^{(i)}_2)=00bb000000110001$. The order of the two input variables actually matter. In the example, the z-value of $(X^{(i)}_2,X^{(i)}_1)$ is $00bb000000110010$. Note that, this z-ordering mapping can actually be modeled by a function with the modulo operator and simple arithmetic operators.

We construct a new sample set as quadratic large as the original set of PerfConf-performance samples by permuting every pair of original samples.  The permutation generates $P_n^2=n\times(n-1)$ samples from the original $n$ samples. On construction, we exploit the above SFC method to map pairs of PerfConf settings into a space with the same dimensions as the number of PerfConfs. It is common practice that inputs are normalized before training machine learning models. Assuming that $X_1,X_2$ are normalized and transformed into the unit interval $[0,1]$, the SFC-based bijection is $h(X_1,X_2)=\overrightarrow{X_{1,2}}$ with $X_1,X_2,\overrightarrow{X_{1,2}}\in [0,1]^d$.

We can generate even more training samples based on historical tuning experiences. Experiences useful for sample generation are comparison-based rules, for example, \emph{increasing the value of PerfConf X leads to a higher performance}. For any given PerfConf setting, we can increase the value of PerfConf $X$ and obtain pairs of PerfConf settings. We can then induce new training samples based on the above sample induction method. As long as the experience-based rule holds, we can generate as many training samples as needed. However, we must be careful of two things. First, the experience-based rule must be correct; otherwise, the model trained on the generated samples would be wrong. Second, we must introduce no data skewness and take samples uniformly distributed in the input space; otherwise, the trained model can be misguiding.\vspace{-6pt}

\subsection{Modeling Comparison as Classification}
\label{sec:classify}%

We can model the comparison-based relations using the machine leaning method of classification, the model of which is called \emph{classifier}. A classification problem is to decide which class a given input belongs to. Given pairs of PerfConfs, we classify their performance comparison results into two classes, i.e., the first better than the second and otherwise. For example, a PerfConf pair $(X_1,X_2)$ is classified into one class if $X_1$ performs better than $X_2$, i.e., when $g(X_1,X_2)=1$; otherwise, it is classified into the other class.

With sample induction $h(X_1,X_2)=\overrightarrow{X_{1,2}}$ as defined in Section~\ref{sec:sampleMapping}, we can transform $g$ of Eq. (\ref{eq:classifier}) into the following function $g'$:\vspace{-6pt}
\begin{equation}
g'(h(X_1,X_2))=
\begin{cases}
1 & \text{if $f(X_1)-f(X_2)>0$},\\
0 & \text{otherwise}.
\end{cases}%
\label{eq:classifierbij}
\end{equation}
where $g(X_1,X_2)$$=$$g'(h(X_1,X_2))$$=$$g'(\overrightarrow{X_{1,2}})$. The input space of $g'$ has the same dimensions as that of $f$, i.e., half the input dimensions of $g$, but with training samples as quadratically many as those for $f$. We can now construct a classifier on the sample set ${(\overrightarrow{X},g'(\overrightarrow{X}))}$ with enough samples.

We might also train a classifier for telling whether one configuration setting is better than the default configuration setting. But this way of constructing a classifier cannot solve the problem of sample scarcity. As our target is to exploit classifier models to solve the tuning problem, our focus is how to use the machine learning model, instead of improving the model. We do not tune the hyper-parameters of the classifier, as this is a problem as difficult as the one that the classifier is trained for. Rather, we bear in mind that the classifier is not precise. We thus design algorithms that could exploit imprecise predictions by such classifier to fulfill tuning-related tasks.

\textbf{Classification vs. ranking.} As related works formulate tuning as an optimization problem, some would think that modeling tuning as a \emph{ranking} problem~\cite{perfRank} would be more natural than as a comparison one. We do not address the tuning problem by ranking models but with classification models for two reasons. First, the input space of configuration tuning generally has continuous dimensions, which would contain in any given range points in a number larger than the total number of natural numbers. As ranking is in fact mapping natural numbers to inputs, this fact indicates ranking is an inadequate way of modeling. Second, configuration tuning is to find the top input(s) in the set, rather than aligning all inputs. While given a ranking model, obtaining any comparison result is straight forward. Given a classification model for comparison, finding the ranking is an NP-hard problem~\cite{orderNPhard}. In other words, the ranking model has incorporated more information than the classification model. That said, like directly predicting performance, performance ranking has also done more than required.\vspace{-6pt}

\section{Tuning with an Imprecise Classifier}
\label{sec:tuning}

With the comparison-based classification model, ClassyTune can compare any pair of PerfConf settings. Since we can now use the trained model as the surrogate of the real system, our goal now becomes to find out the best one in a sufficiently large set of $N$ PerfConf settings.

\textbf{Strawman.} One naive solution is to sample $N$ PerfConf settings and use the classifier to compare every pair of them. In order to find the optimal setting, $N$ must be sufficiently large to cover the whole space of PerfConfs. Unfortunately, pairing every two of the $N$ PerfConf settings would lead to a set with a daunting size of $C_N^2$. Even though the classifier can predict in a sufficiently short time, this processing time would \emph{add up to a long duration}. Worse still, as the classifier is not a hundred percent correct, \emph{some results would be contradicting}, making it impossible to deduce the real optimal.

\textbf{A better strawman.} An alternative solution is to do a binary search among the huge set of $N$ PerfConf settings. In each comparison, i.e., each prediction by the classifier, the winning PerfConf setting is kept for the next round of comparison, while the other one is discarded directly. After $\log_2 N$ rounds of binary comparisons, we will finally reach the last pair of winning PerfConf settings. And, the final winner will be the optimal. However, as we have mentioned, the classifier is not a hundred percent correct; thus, \emph{the actual optimal setting might have been discarded because it loses in just one false comparison.}

\textbf{Our solution.} ClassyTune takes a systematic approach towards tuning. Rather than trying to improve the precision of the model, ClassyTune recognizes that the trained model can only make a large portion of predictions correct. It exploits this fact and finds the top setting in best effort through three phases, i.e., finding a list of good PerfConf settings, locating promising areas with optimal settings and searching for the optimal setting.\vspace{-6pt}

\subsection{Finding Better PerfConf Settings}

ClassyTune does not compare every pair of PerfConf settings. Rather, in the training phase, it keeps the best PerfConf setting in the training set along with the trained model. When given the large set of $N$ PerfConf settings, ClassyTune uses the trained model to compare each of the $N$ settings with the best PerfConf setting in the training set. This list of settings that win in the comparisons are kept. Even though the trained model might not be completely correct in these comparisons, it is very likely that many of these winning settings are ones  better than the best PerfConf setting in the training set.

We take \emph{a list of winning settings} output by the imprecise classifier. We do not keep the single PerfConf setting that wins the most comparisons, contrasting the way that BO with the GP prior takes one optimal setting at each step. Given the same imprecision rate, finding a list of winning settings reduces the probability that we find no PerfConf setting better than the best one in the training set.

Furthermore, we do not directly output this list of winning PerfConf settings as optimal ones. Rather, we use them to locate some promising areas for finding the real optimal setting. The reasons include: 1) as the model is not a hundred percent accurate, some of the winning settings might not even be good settings; and, 2) the space of PerfConfs is too large such that the $N$ settings might not be representative enough for finding the optimal one.

\subsection{Clustering to Locate Promising Subspaces}

In fact, we believe good settings are close to each other and possibly locate at a few promising areas. Generally, the optimal PerfConf setting is surrounded by good settings that are better than many others. Likewise, the areas where many good settings locate are promising places that the optimal setting might be found. We denote such areas as the \emph{promising} subspaces.

For the set of winning PerfConf settings, ClassyTune uses the clustering algorithm of KMeans to find out where the good PerfConf settings cluster. To determine the number of promising areas, i.e., the number of clusters, we exploit the elbow criterion~\cite{elbowclusternum} to find a best number $k$ for clustering. We then run the KMeans algorithm to cluster the winning PerfConf settings into $k$ clusters, whose centers are then computed. The promising subspaces are located around these centers.

\subsection{Searching for the Best}

Now, we have the centers of the promising subspaces. We have not yet set their boundaries. We set the boundaries of the promising subspaces based on the PerfConf settings that we have already evaluated. As we know that none of the evaluated settings is expected to be better than the list of winning settings, we should not consider those settings lying farther from any center than an evaluated setting that is closer to the center than other evaluated settings. Hence, for each center, we find at each dimension its closest neighbor in the set of evaluated settings; and, the value of this neighboring setting is used as the boundary for this dimension by the center. After finding for each dimension of all centers, we bound all promising subspaces.

Within the specified number of tuning tests, we then sample in the promising subspaces so that a good coverage of the areas is guaranteed~\cite{bestconfig}. These sampled PerfConf settings are then evaluated in the system to decide which exactly is the best. The final best will be output as the suggested setting for an optimal performance.\vspace{-6pt}

\section{The ClassyTune System}
\label{sec:implement}

\begin{figure}[!t]
      \centering
      \includegraphics[width=0.46\textwidth]{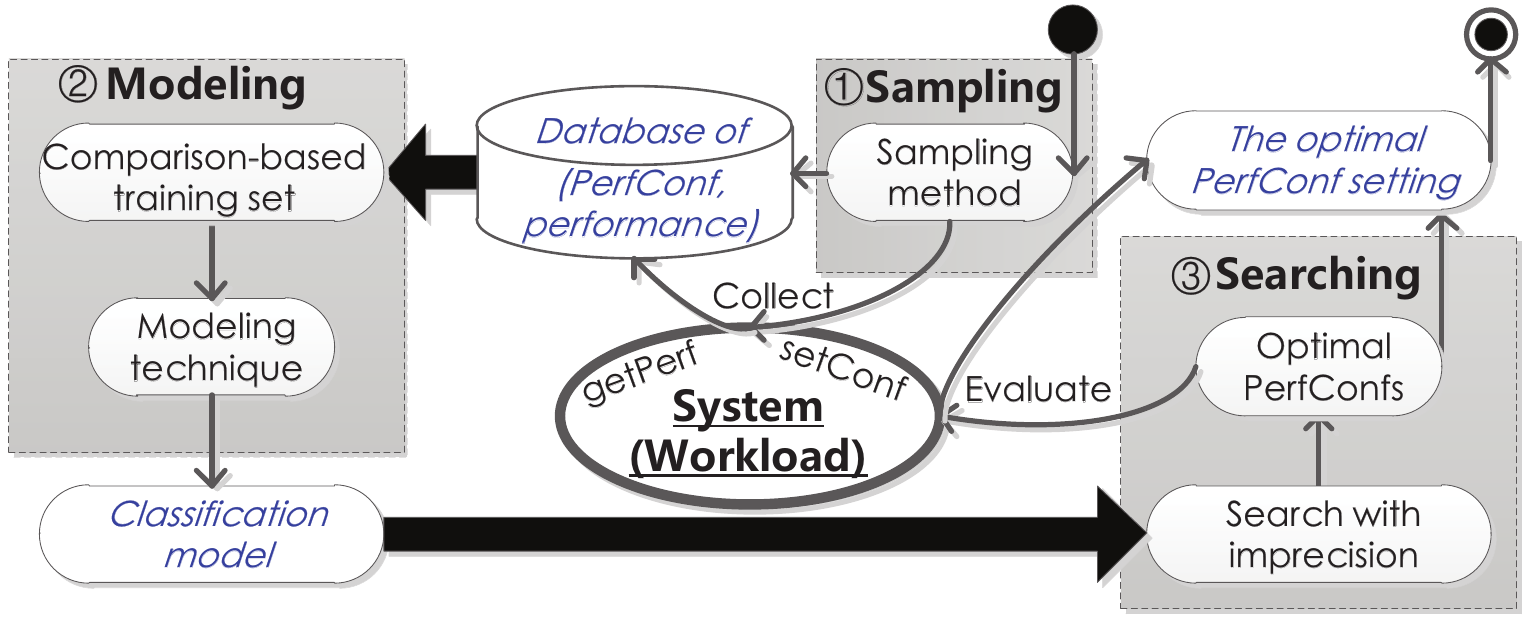}\vspace{-6pt}
      \caption{ClassyTune: the architecture \& the tuning process.}\vspace{-12pt}
      \label{fig:ClassyTune}
\end{figure}

The overall architecture of ClassyTune is illustrated in Figure~\ref{fig:ClassyTune}. Like BestConfig~\cite{bestconfig} and Ottertune~\cite{ottertune}, ClassyTune only needs the users to provide a list of PerfConfs along with their valid ranges, and scripts to set PerfConf values/get system performances for tuning a new system and its application workload. ClassyTune has three main components, i.e., sampling, modeling and searching. These components interact through data flows, thus they can locate on one same server or multiple servers. The results of sampling and modeling are produced as the intermediate outputs for reuse in following tasks. The two intermediate outputs are the database of PerfConf-performance samples and the classifier model. The final output of the tuning process is the best PerfConf setting found within the given number of tuning tests.
\subsection{Main Components}%
\label{sec:components}%

\textbf{Sampling.} Different from common machine learning tasks, configuration tuning allows the learning process to freely choose the points to sample in the input space. As all values in the range are valid for a dimension, sample values on each dimension should spread across the corresponding range so that the underlying relations impacting comparison results should be represented and learned. According our practical experience, we find the latin hypercube sampling (LHS) method~\cite{lhs} used in ClassyTune very effective and to the purpose. It can (1) uniformly cover the whole range on each dimension and (2) sample a given number of points. In comparison, uniform random sampling does not necessarily cover the whole range, while grid sampling might not be able to sample for a required number of points. Other sampling methods that satisfy the two properties like LHS can also be used with ClassyTune. The output of the sampling phase is a database of PerfConf-performance samples.

\textbf{Modeling.} ClassyTune exploits the database of PerfConf-performance samples to construct news samples for training the comparison-based model. ClassyTune tries different classification methods to train the comparison-based model ($\S$\ref{sec:classify}). This comparison-based modeling enables ClassyTune to discover the latent principle underlying how changing an input leads to the change of performance. In comparison to direct performance modeling, ClassyTune simplifies the task by not requiring the precise prediction on the performance numbers. Rather, ClassyTune only cares about the relative relations of resulting performances. The constructed model is output as an intermediate result of the tuning process. It can be exploited later by other analysis tasks for the system.

\textbf{Searching.} Based on the classifier, we search the configuration space thoroughly for a set of best points. The classifier is used to decide whether a configuration setting is superior to any other configuration. \emph{This prediction takes much less time than actually evaluating a configuration setting for the performance.} Like other model-based tuning solutions~\cite{metis,ottertune,cartModel}, ClassyTune exploits the trained model as a surrogate. Different from some Bayesian-optimization based solutions that explicitly solve an optimization equation, ClassyTune takes a systematic approach to optimization, adopting a three-phase searching process. The found candidate settings are evaluated in the system for verification.

\subsection{The Tuning Algorithm}
\label{sec:algo}

The whole tuning process of ClassyTune is implemented as illustrated in Algorithm~\ref{algo:opt}. Given a set of PerfConf-performance samples as input, we first induce a new sample set for training a binary classifier (Line~1). Then, we find the best PerfConf setting in the original sample set (Line~2). Using the best PerfConf setting in the training set as the pivot, ClassyTune compare each of the $N$ PerfConf settings with this pivot (Line~3-5). All the winning settings are put in a winner set (Line~6-7). Second, ClassyTune proposes to enclose the areas where the winner settings cluster in (Line~8-9). These areas are the promising subspaces where optimal settings might locate. Even though the classifier might have mispredicted some winners, the location of the promising subspaces might be shifted a little bit but would not be completely missed. Third, to actually find the optimal settings, ClassyTune proposes to resample in these subspaces and evaluate the sampled PerfConf settings in the system (Line~10-11). The best setting will be output as the solution (Line~12-14).
\setlength{\textfloatsep}{0.6\baselineskip}
\begin{algorithm}[!t]
\footnotesize
\caption{ClassyTune: classification-based tuning.}%
\label{algo:opt}%
\KwIn{$X, y$\tcp*{\scriptsize PerfConf settings,performance}}
\KwIn{$m$\tcp*{\scriptsize \# of validating PerfConf settings}}
\KwOut{$bestX$\tcp*{\scriptsize the optimal PerfConf setting}}
\tcc{\scriptsize induce samples, train classifier}
$clf$=\texttt{\footnotesize FIT}(\texttt{\footnotesize SET\_INDUCE}($X,y$))\;
$idxMax$=\texttt{\footnotesize ARGSORT}($y$)[-1]\tcp*{\scriptsize index of best $y$}
\tcc{\scriptsize sample many points in the space}
$S\leftarrow\{X\}$ \tcp*{\scriptsize $\mid S\mid>1000\times DIM$($X$)}
$Xp$=\texttt{\footnotesize PAIR\_INDUCE}($S$,$X$[$idxMax$])\;
$Yp$=$clf$.\texttt{\footnotesize PREDICT}($Xp$)\;
\tcc{\scriptsize points better than $X$[$idxMax$]}
$idxList$=\texttt{\footnotesize IDX\_WHERE}($Yp$,$yp_i>0$)\;
$X_s$=$S$[$idxList$]\;
\tcc{\scriptsize compute best \# of clusters}
$k$=\texttt{\footnotesize BEST\_CLUSTER\_NUM}($X_s$)\;
\tcc{\scriptsize cluster points to promising subspaces}
$C$=\texttt{\footnotesize KMEANS\_FIT\_AND\_GET\_CENTERS}($k$,$X_s$)\;
\tcc{\scriptsize sampling in promising subspaces by \texttt{\footnotesize LHS}}
$X\_candidates\leftarrow$\texttt{\footnotesize LHS}($C$, $m/k$)\;
$y\_candidates\leftarrow$\texttt{\footnotesize EVALUATE}($X\_candidates$)\;
$idxMax$=\texttt{\footnotesize ARGSORT}($y\_candidates$)[-1]\;
$bestX\leftarrow X\_candidates$[$idxMax$]\;
return $bestX$\;
\end{algorithm}
\subsection{Implementation}
\label{sec:sysdetails}

\textbf{Data types for sample representation.} One could notice that, we need to use data types with higher precision to represent the induced samples. In our implementation, we use the \emph{double} data type to represent the induced sample values and the \emph{float} for the original ones. However, the lengthy tail of a decimal is very likely to lose its significance in the model training process. Thanks to the sparsity of samples, it is rare that the induced inputs get collapsed with the original ones. The disadvantage of the induction is that the latent relations between configuration pairs could become even further profound. However, as we have mentioned in Section~\ref{sec:sampleMapping}, the sample induction can actually be modeled as a function of modulo and other simple arithmetic operators. Luckily, as demonstrated by many real-world applications, some classification algorithms can represent highly complex input data~\cite{evalClassifier1}.

\textbf{Other implementation details.} We implement ClassyTune using Python and R, with only about 2000 lines of code. The interactions with the system under tune are implemented through shell scripts. ClassyTune maximizes a scalar performance metric. The scalar performance metric can be defined and specified through some utility function~\cite{bestconfig}, with user-concerned performance goals as inputs.
\begin{table}[!t]
  \caption{The Evaluated Systems and Variables}\vspace{-6pt}%
  \label{tbl:suts}%
  \centering
  \begin{tabular}{p{1.2cm}lp{0.5cm}c}
\toprule[1.2pt]
{\small \textbf{System}} & {\small \textbf{Description}} & {\small \textbf{Lang.}} & \textbf{\small Workloads}\\
  \midrule[0.8pt]
  \textbf{HDFS} & Dist. filesystem  & Java & PageRank,\\
  \textbf{YARN} & Dist. processing  & Java & Join,\\
  \textbf{Hive} & Data analytics  & Java & KMeans\\
  \midrule[0.2pt]
  \textbf{Spark} & Data processing  & Scala &PageRank,TeraSort,KMeans\\
  \midrule[0.2pt]
  \textbf{MySQL} & DB server  & C++ & readOnly,readWrite,TPC-C\\
  \midrule[0.2pt]
  \textbf{PostgreSQL} & DB server  & C & readOnly,readWrite,TPC-C\\
  \midrule[0.2pt]
  \textbf{Cassandra} & NoSQL DB & Java & readWrite(YCSB-a)\\
  \midrule[0.2pt]
  \textbf{Tomcat} & Web server & Java & Web exploration\\
\bottomrule[1.2pt]
\end{tabular}
\end{table}
\section{Evaluation}%
\label{sec:eval}%

\subsection{Experimental Settings}

We evaluate ClassyTune over 7 cloud systems that are implemented in different languages. They have supported a variety of applications. These systems are listed in Table~\ref{tbl:suts}. To provide an example of tuning co-deployed cloud systems, we tune Hive and Hadoop together for offline data analytical workloads. We choose these systems in accordance with related works~\cite{smartconf,bestconfig,ottertune} for an easy comparison. We believe our choice should be representative for a large number of cloud systems.

We choose 14 application workloads following the choice of related works~\cite{smartconf,bestconfig,ottertune}, as listed in Table~\ref{tbl:suts}. The cases of Tomcat and Cassandra are relatively simple as compared to other systems, so only the workloads of Web exploration and read-write are chosen respectively. The other systems are evaluated on three typical workloads. The distributed processing systems of Spark and Hive plus Hadoop are evaluated under analytical and machine learning workloads, generated by the HiBench benchmark. The transactional (readWrite) and readOnly workloads for databases are generated by the SysBench benchmark. We also include the TPC-C workload, the current industrial standard for evaluating the performance of OLTP systems.

For each system, we choose 10 influential PerfConfs for tuning, unless mentioned otherwise. Related works taking the model-based approach typically use a similar number of parameters, around 7 to 16 and with 8 achieving the best on tuning with fixed parameters~\cite{metis,ottertune}. We choose the PerfConfs to tune in accordance with related works. These PerfConfs control various aspects of systems, including but not limited to network, CPU, memory, storage, indexing, caching and buffering.

Performance metrics are application-specific. We adopt the performance metrics commonly used for the evaluated workloads. While workloads on  Spark and Hive plus Hadoop are tuned for a shorter processing time (or task duration), workloads on the other systems are tuned for higher throughputs.

Our experimental platform consists of 12 servers. Each server has two 12-core Intel Xeon E5620 CPU with 32GB RAM. CentOS 6.5 and JVM 1.7 are installed. For each evaluation, one server is used to generate workloads. Standalone SUTs are run on one server, while distributed SUTs are hosted by four servers.\vspace{-6pt}

\subsection{Selecting the Classification Model}
\label{sec:modelselect}%

We empirically study which classifier is best to be used with ClassyTune.  There exist many machine learning methods to model the comparison relations, e.g., logistic regression (LR for short), decision tree (DT), supported vector machine (SVM), neural networks (NN) and XGBoost. While the former three are the classic methods for binary classification, neural networks have been applied to many real applications and make significant progress in applications to scenarios with big data. XGBoost (XGB for short) is in the algorithm family of gradient-boosted trees~\cite{gbt}, which have been shown to be among the best classifiers ~\cite{evalClassifier1}. In binary classification problems with small data, algorithms from the families of gradient-boosted trees are on the top among all. XGBoost has been used to achieve state-of-the-art results on many machine learning challenges.

\begin{figure}[!t]
      \centering
      \includegraphics[width=0.5\textwidth,height=83pt]{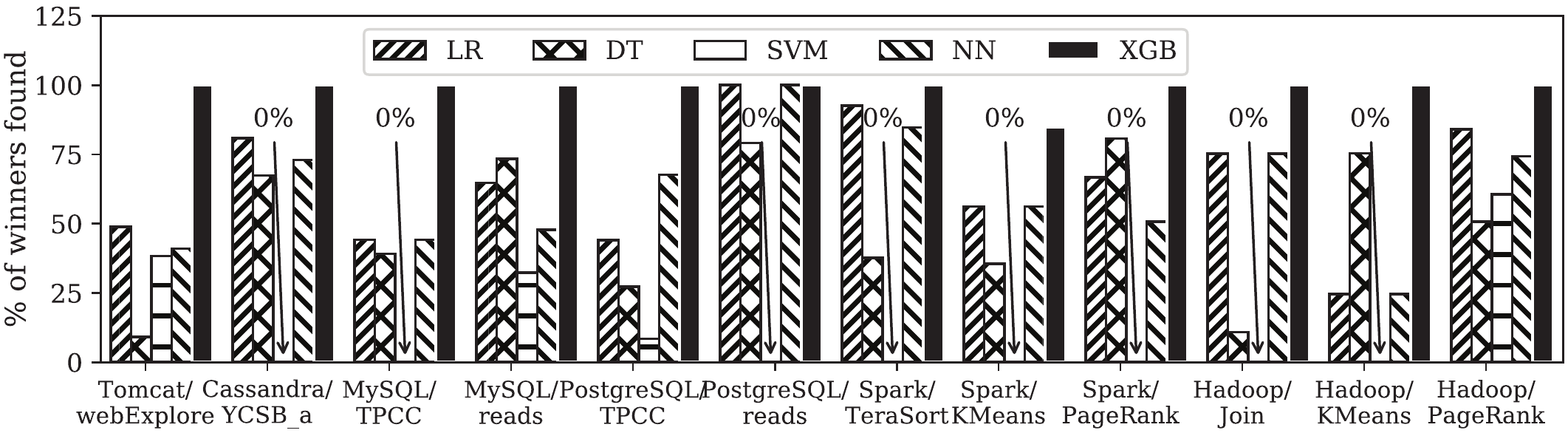}\vspace{-6pt}
      \caption{Percentage of winning settings found by different classifiers: XGB outperms all the other classifiers, while the kernel method SVM, exploiting covariance functions, fails in most cases.}
      \label{fig:classifier} 
\end{figure}
In the comparison-based tuning, the key to success is to recognize the whole set of PerfConf settings that are better than and winning a given one. We evaluate the above five classifiers to see how they can recognize the winning settings. We let each classifier to be trained on a set of 50 original samples and tested on 20 samples. The 20 samples have performances higher than the best sample in the training set. We evaluate to see how many among the 20 samples can be recognized by a trained classifier. The results are plotted in Figure~\ref{fig:classifier}. From Figure~\ref{fig:classifier}, we can see that XGBoost can almost find all the winning settings for all systems. Therefore, we choose XGBoost as our classifier model in ClassyTune.\vspace{-6pt}

\subsection{Tuning Efficacy}

\textbf{Comparing to performance-prediction based tuning.} We have tried predicting winning settings using \emph{regression} models on the same set of original samples as in Figure~\ref{fig:classifier}. We use the decision tree based regression model, which is shown to perform best in predicting system performances~\cite{cartModel}. But the model trained on the same sample set \emph{fails to find out any} of the winning samples. This again proves the validity of taking a comparison-based approach.\vspace{3pt}

\textbf{Compared to other auto-tuning methods}. To demonstrate the tuning efficacy of ClassyTune, we compare ClassyTune with two state-of-the-art tuning approaches, i.e., the search-based approach~\cite{bestconfig} and the Gaussian-process (GP) based Bayesian optimization (BO) approach~\cite{metis,ottertune}. Besides,\emph{ these two approaches are the few auto-tuning proposals that work on a limited number of samples}. We do not compare with approaches based on control theory~\cite{smartconf} or reinforcement learning~\cite{capes} because they are only applicable to a handful of configuration parameters. We exploit the open-source implementation BestConfig\footnote{\scriptsize http://github.com/zhuyuqing/bestconf/} for the evaluation of the search-based approach. As no open-source implementation can be found for the GP-based BO tuning approach~\cite{metis,ottertune}, we implemented it exploiting the Python package of GP-based BO implementation\footnote{\scriptsize https://github.com/thuijskens/bayesian-optimization}.
\begin{figure}[!t]
\vspace{-9pt}
\centering
 \subfloat[Throughputs of Web server, NoSQL database, and databases.]{
       \includegraphics[width=.23\textwidth]{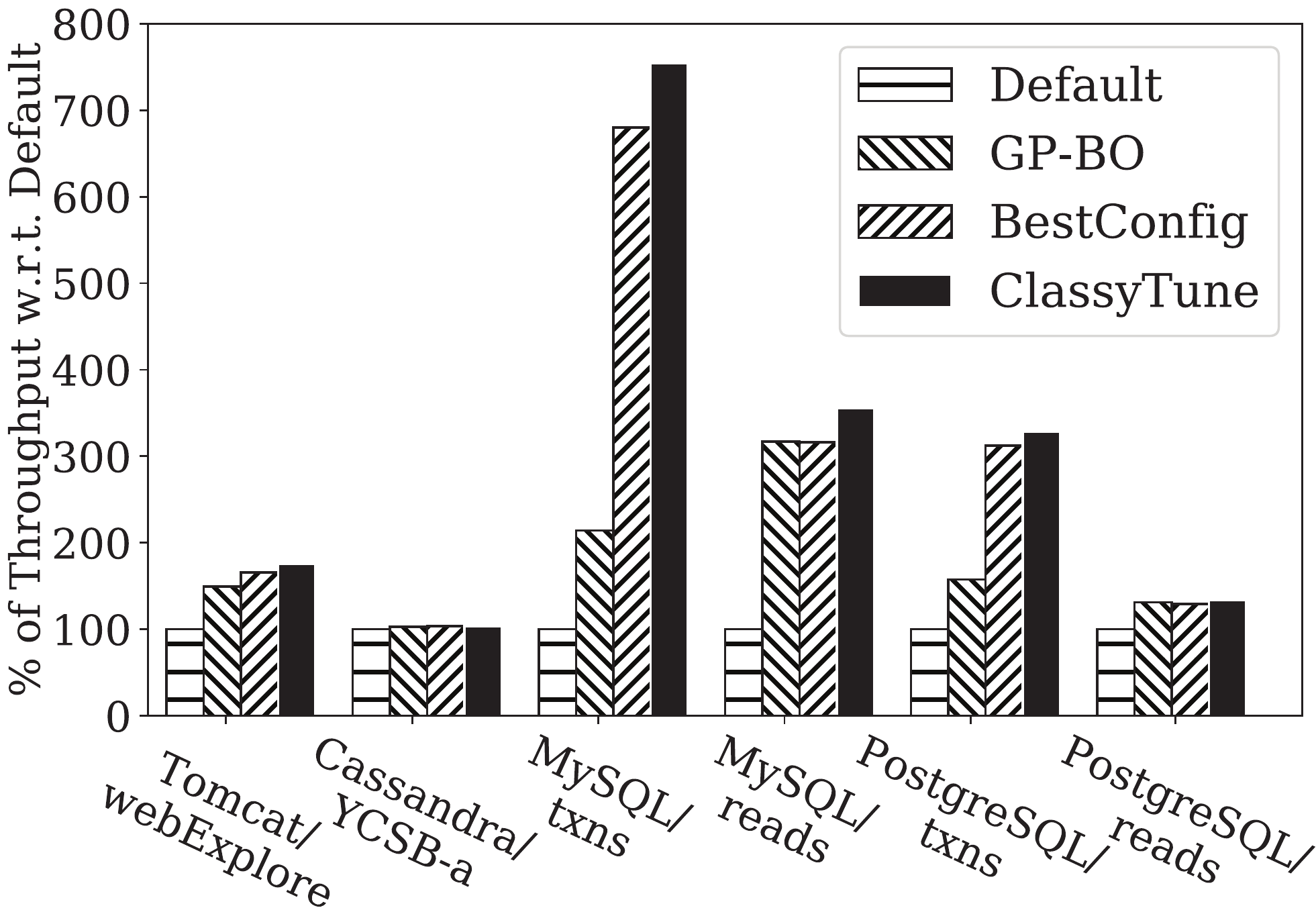}%
        \label{fig:efficacy:throughput} 
    }\vspace{-3pt}
  \subfloat[Running times of Spark and Hadoop jobs.]{
        \includegraphics[width=.23\textwidth]{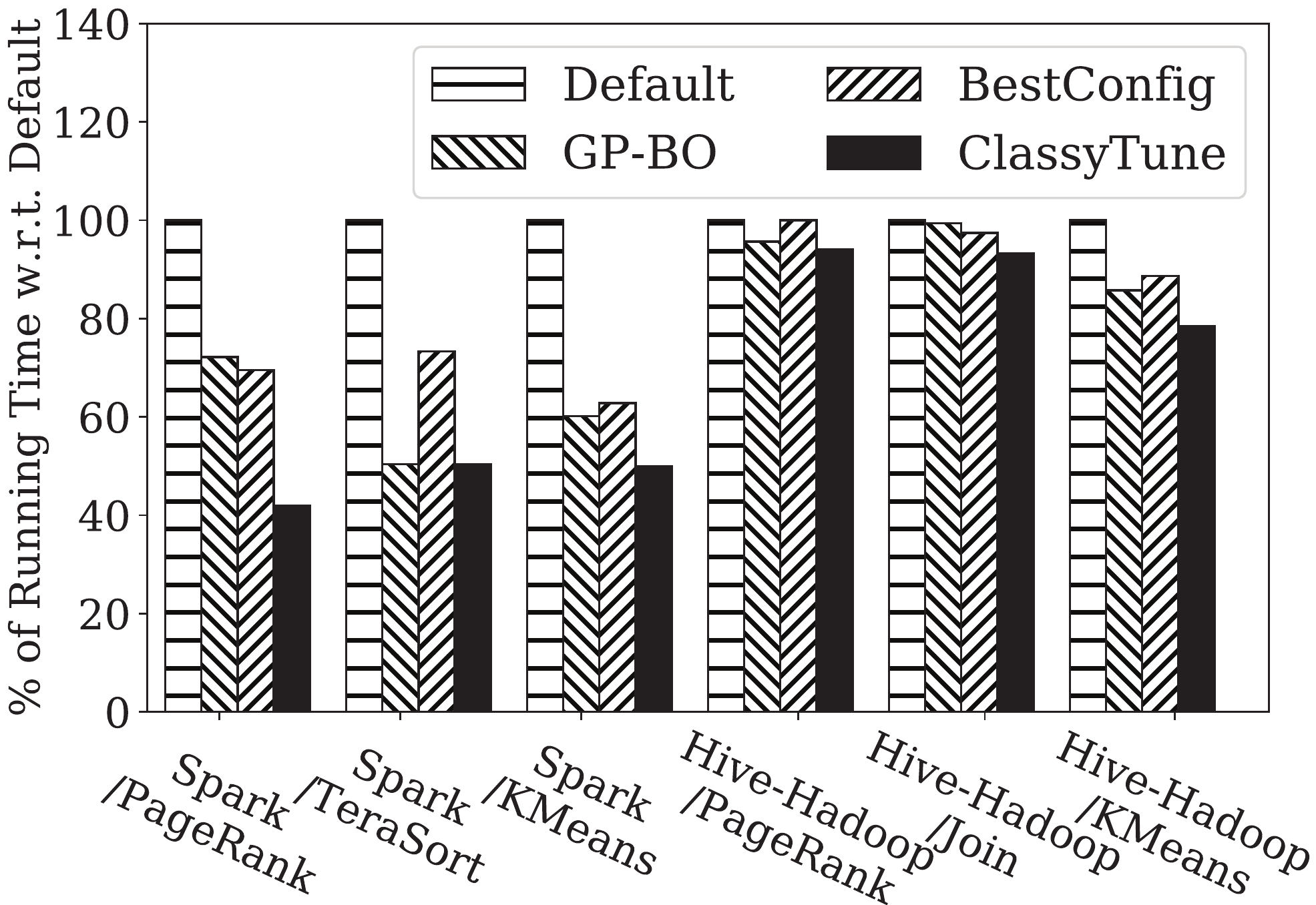}%
        \label{fig:efficacy:duration} 
    }\vspace{-3pt}
     \caption{ClassyTune/BestConfig/GP-based BO(GP-BO) improving performances over those under default settings.}
     \label{fig:efficacy}
\end{figure}
\begin{figure*}[t]
\centering
\hspace{3pt}
 \begin{minipage}{0.3\textwidth}
      \centering
\vspace{6pt}
     \includegraphics[width=\textwidth,height=97pt]{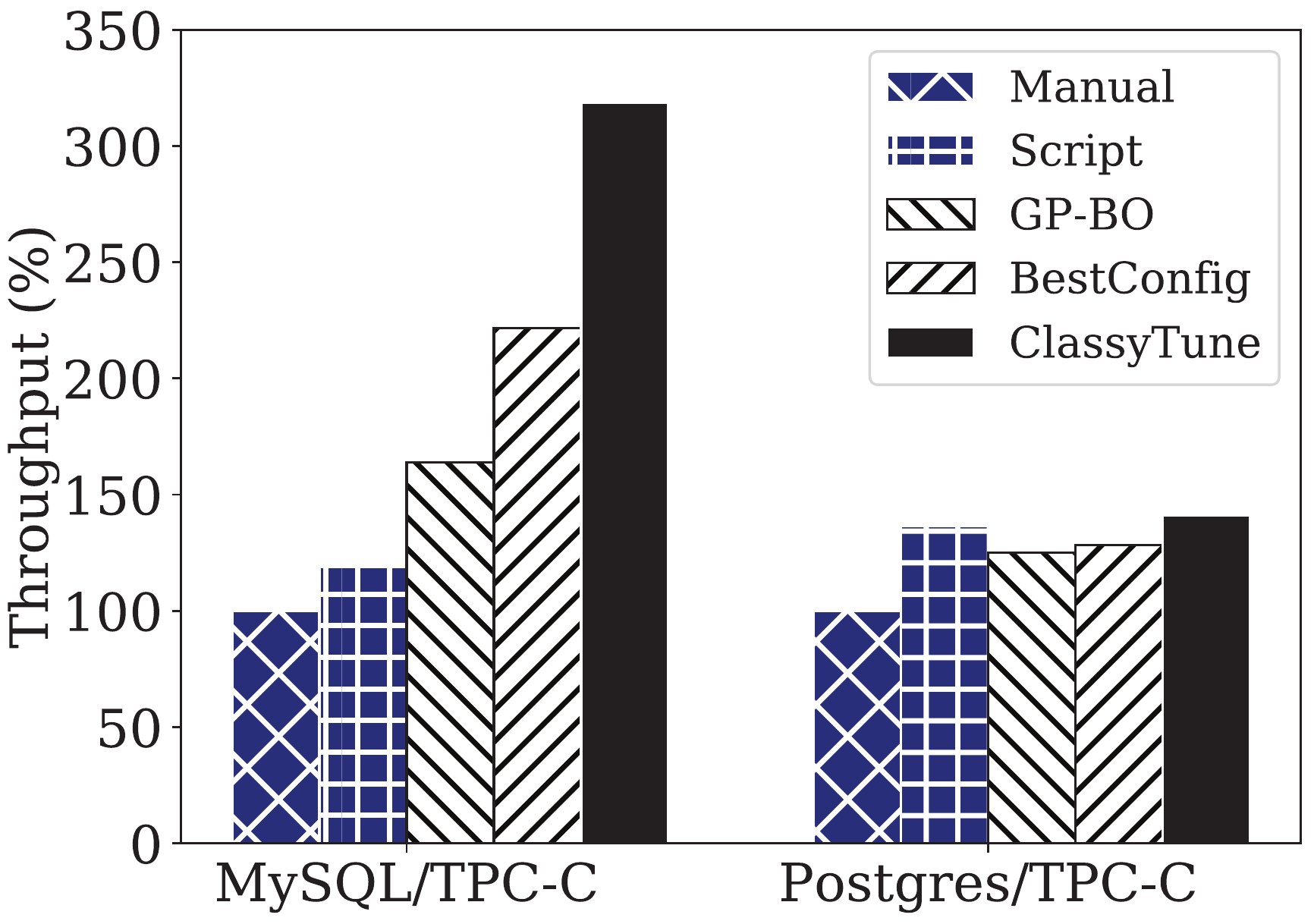}
     \caption{Auto-tuning compared to manual tuning: databases/TPC-C.}\vspace{-12pt}
     \label{fig:dbs}
  \end{minipage}
  \hspace{6pt}
 \begin{minipage}{0.64\textwidth}
    \subfloat[PageRank on Spark]{
    \centering%
       \includegraphics[width=.43\textwidth,height=85pt]{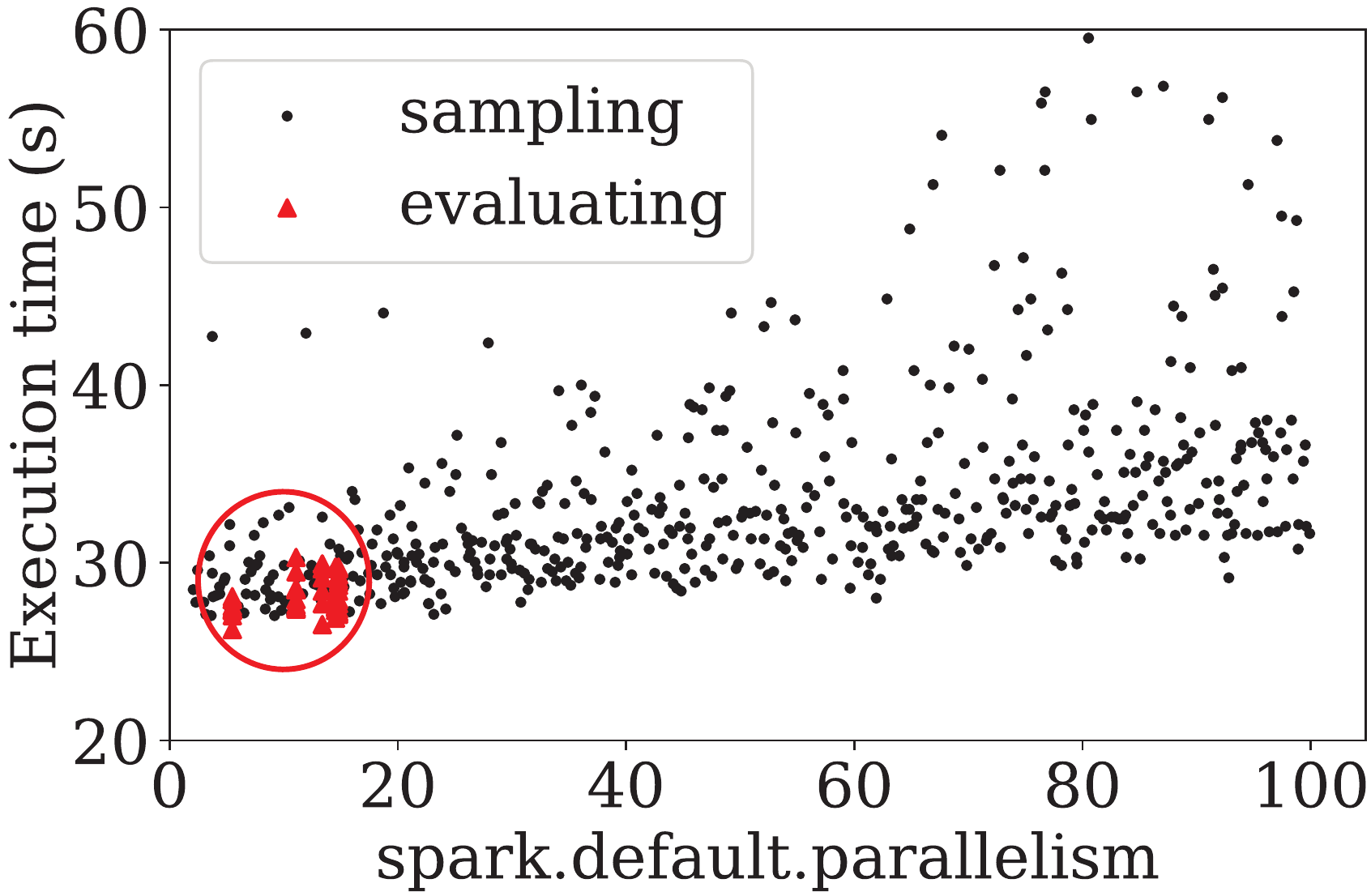}%
        \label{fig:cluster:spark} 
    }
    \hspace{3pt}
    \subfloat[PageRank on Hive-Hadoop]{
    \centering%
        \includegraphics[width=.43\textwidth,height=85pt]{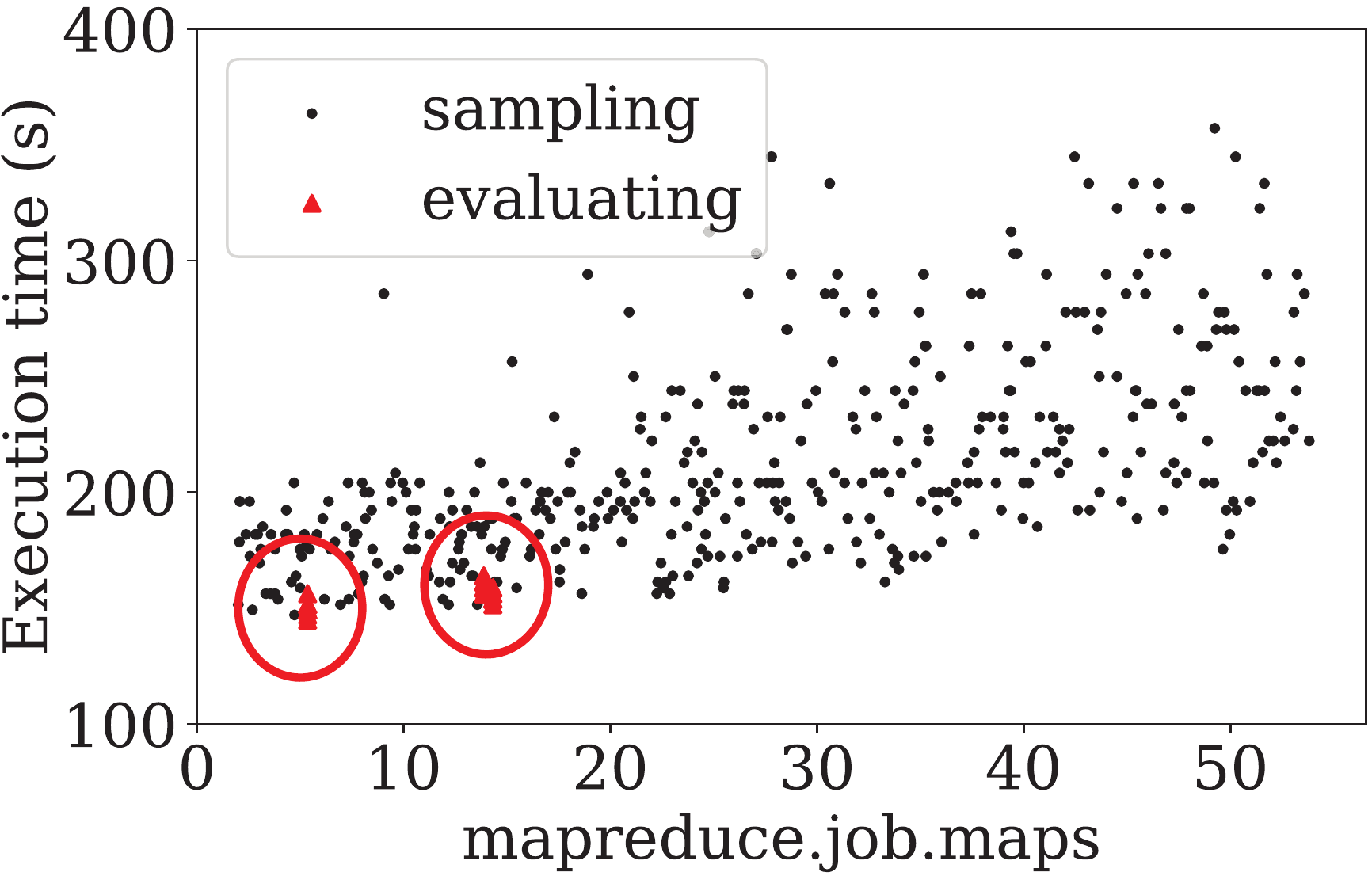}%
        \label{fig:cluster:hadoop} 
    }
    \caption{Promising subspaces (bounded by circles) with optimal settings (i.e., evaluating points) as located by ClassyTune.}\vspace{-12pt}
    \label{fig:cluster}
  \end{minipage}
\end{figure*}

For each combination of tuning solutions, systems and workloads, we run the tuning experiment for three times and report the average performance improvement. In each tuning experiment, we tune within 100 tests, as following the evaluation methodology of related works~\cite{bestconfig}.

Figure~\ref{fig:efficacy} shows that ClassyTune can find configurations better than and occasionally as good as those output by the two state-of-the-art solutions. ClassyTune can improve throughputs to as much as about  $8\times$ of that under the default setting, and decrease execution times to as much as about $1/4$. Specifically, it has improved the throughputs of Tomcat by 76\%, Cassandra by 4\%, MySQL/transactions by 654\%, MySQL/reads by 256\%, PostgreSQL/transactions by 228\% and PostgreSQL/reads by 33\%. It reduces the execution time by 58\% for Spark/PageRank, 72\% for Spark/TeraSort,  50\% for Spark/ KMeans, 6\% for Hive-Hadoop/PageRank, 7\% for Hive-Hadoop/Join and 22\% for Hive-Hadoop/KMeans.

Even for the complex co-deployed system of Hive-Hadoop, ClassyTune can still improve the performance by reducing as much as 22\% execution time of the KMeans workload. In comparison, the search-based method and the GP-based BO method cannot tune such a complex system to a performance as good as ClassyTune.

ClassyTune tunes several systems to a performance much higher than the state-of-the-art solutions, e.g., Spark/PageRank and MySQL/txns in Figure~\ref{fig:efficacy}. For other systems, ClassyTune can only win the state-of-the-art solutions by a small percentage. A system can in no way be tuned as well as one would wish by only changing PerfConf settings. There is an upper bound on the performance that tuning PerfConf settings can improve, although this bound can hardly be figured out for the high-dimensional continuous space of PerfConfs. The performances that ClassyTune has tuned to are the best we have found for the corresponding combinations of systems, workloads and environments. We have tried testing each combination over thousands of different PerfConf settings, but we never find one setting better than the one suggested by ClassyTune.\vspace{3pt}

\textbf{Compared to manual and expert tuning.} To further demonstrate the effectiveness of ClassyTune, we also evaluate the performances tuned by ClassyTune towards those tuned manually or by expert knowledge. We experiment with databases under the TPC-C workload. To enable the comparison, we adopt the setting as suggested by the Internet and related works~\cite{ottertune} for the manual setting. Before automatic tuning appears, a common way for tuning databases is to use scripts that are written by experts based on their knowledge and expertise. We exploit two tuning scripts for MySQL\footnote{\scriptsize https://launchpad.net/mysql-tuning-primer} and PostgreSQL\footnote{\scriptsize http://pgfoundry.org/projects/pgtune/} respectively. These scripts are also evaluated in a related work~\cite{ottertune}. We also demonstrate the tuning results of GP-based BO and BestConfig. Figure~\ref{fig:dbs} presents the results.

ClassyTune can improve the system performance to about $3.2\times$ of that under the manually tuned configuration. In fact, human beings can hardly capture fully the characteristics of complicated workloads, thus auto-tuning methods find PerfConf settings with better performances than those under manual-tuned and script-tuned PerfConf settings. And, the latent relations between PerfConfs and performances are better captured if modeled in the way of ClassyTune than if modeled in the way of GP-based BO. Therefore, ClassyTune has an advantage in both database cases, while the BO-based and the search-based approaches perform slightly worse than the script-based approach on tuning PostgreSQL. We believe that the number of samples is an influential factor. ClassyTune acquires its advantage from the comparison-based modeling.\vspace{-6pt}

\subsection{Understanding Comparison-Based Tuning Process}

\textbf{Have winning PerfConf settings been recognized?} We measure to see whether ClassyTune can correctly differentiating all PerfConf settings better than a given one. As plotted in Figure~\ref{fig:classifier}, we can see that the classifier model can almost perfectly identify the list of winning PerfConf settings when only 50 samples are provided. This fact supports our design choice in locating promising subspaces by clustering these winning PerfConf settings.\vspace{3pt}

\textbf{Are promising subspaces located?} We examine whether ClassyTune actually locates the promising subspaces. To better view the PerfConf-performance relations, we run a tuning experiment with 1000 tests for Spark/PageRank and Hive-Hadoop/PageRank respectively. We select the most influential PerfConf \texttt{\small spark.default.parallelism} for Spark and \texttt{\small mapreduce.job.maps} for Hive-Hadoop. We plot all the sampled points in the sampling phase and the evaluated points in the searching phase. The results are shown in Figure~\ref{fig:cluster}. For both systems, the evaluated points are clustering in its space, which is circled out. And, the clusters are having short execution times, i.e., higher performances, than other sampled points. In other words, ClassyTune has successfully located the promising subspaces and recognized a set of good settings.\vspace{3pt}

\textbf{Imprecision is alleviated by the systematic approach.} We further verify the impacts of classifiers' imprecision on tuning. We choose to evaluate on Tomcat-webExplore and PostgreSQL-reads because classifiers display the most difference in the former and the least in the latter in Figure~\ref{fig:classifier}.

XGB, DT and LR improve the performances to 1.76, 1.71 and 1.73 respectively for tuning Tomcat/WebExplore, while they improve to 1.33, 1.25 and 1.24 respectively for PosgreSQL/reads. We can find that the differences between the improved performances are not as much as those between the percentage of winning settings found.

In fact, the tuning results of ClassyTune do not solely rely on the precision of the classifier. Rather, after the classifier pins down the promising areas, we take a systematic approach by resampling in the areas using the LHS method. This result leads us to think that, while exploiting machine learning models are beneficial, taking a systematic approach to the goal will also help to reduce the effect brought about by the imprecision of machine learning models.\vspace{-6pt}

\subsection{Sample Induction Method}

\begin{figure}[!t]
      \centering
      \includegraphics[width=0.47\textwidth,height=85pt]{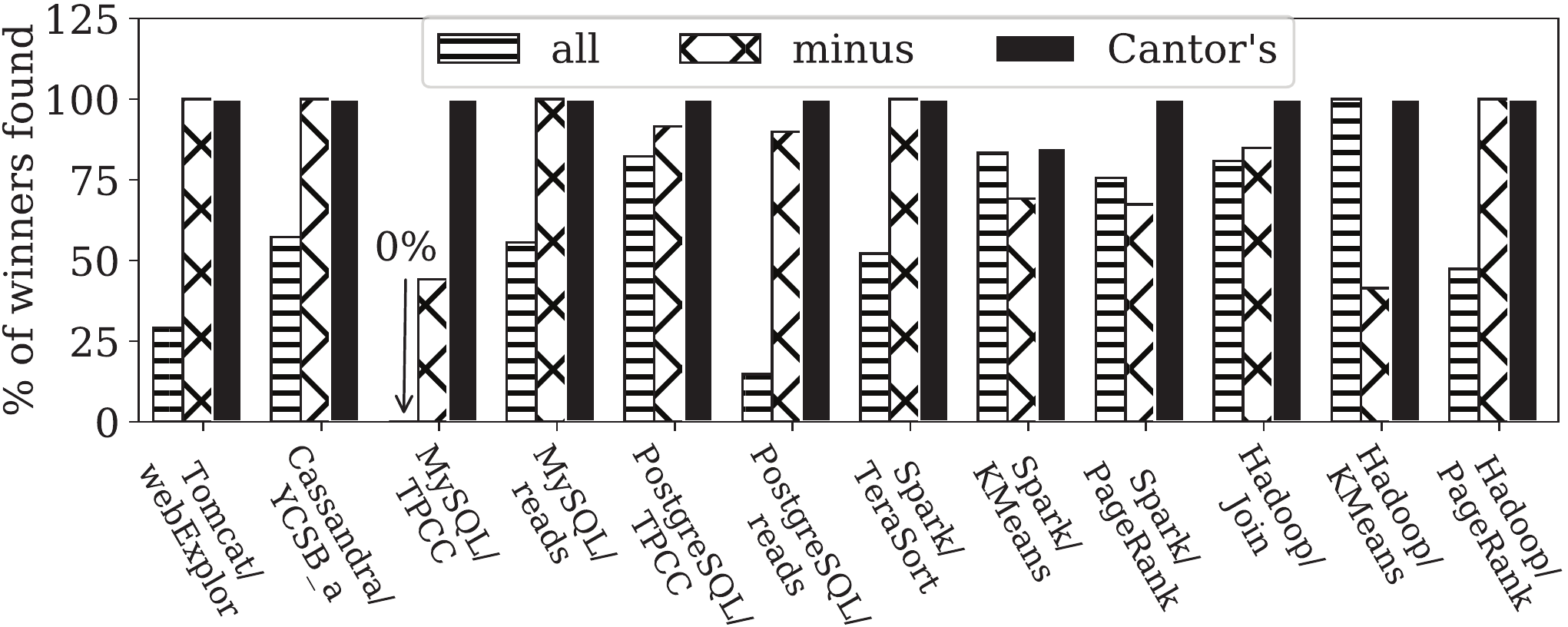}\vspace{-6pt}
      \caption{Percentage of winning settings found: sample induction based on Cantor's proof outperforms others.}
      \label{fig:sampleinduce} 
\end{figure}

We evaluate whether the bijection-based sample induction actually performs better than the simple way of directly taking the difference (i.e., using the minus operation). We also compare our sample induction method with the direct concatenation of two PerfConf settings. We evaluate the three methods on the percentage of winning settings they can find. In the experiments, we use the XGBoost classifier for all three sample induction methods. Results are illustrated in Figure~\ref{fig:sampleinduce}.

Our sample induction method based on the Cantor's proof performs the best for all systems. As we have mentioned in Section~\ref{sec:sampleMapping}, this sample induction method can be modeled as a function of modulo and simple arithmetic operators, although it is seemingly complicated. On the one hand, functions with modulo and simple arithmetic operators can easily be learned by common machine learning algorithms~\cite{learningTheory}. On the other hand, our sample induction method feeds the model with the real independent factors, i.e., PerfConfs. In comparison, the concatenation method mixes independent factors with correlated factors, increasing the input dimension simultaneously. And, the difference method performs worse than our method because the difference computation can lead to collision of mappings.

\subsection{High-dimensional Tuning}

We demonstrate ClassyTune' advantages for tuning in a high-dimensional input space. We choose a tuning space with $30$ PerfConfs and constrain the tuning within $100$ tests. We compare ClassyTune to the two state-of-the-art auto-tuning methods, i.e., the search-based~\cite{bestconfig} and the GP-based BO~\cite{metis,ottertune} approach. Manual tuning is not applicable to high-dimensional tuning because it is very difficult for human beings to comprehend relations in a high-dimensional space, if not impossible~\cite{asilomar}. Script-based tuning is also based on human experiences, making it inapplicable to high-dimensional tuning either. We tune for MySQL and PostgreSQL under the TPC-C workload respectively.

The tuning results are presented in Figure~\ref{fig:hiditune:perf}. First, increasing the dimension leads to a larger input space with possibly even better results, e.g., for MySQL/TPC-C. The performance improvements are higher than those in a 10-dimensional input space, as demonstrated in Figure~\ref{fig:efficacy}. ClassyTune outperforms the other auto-tuning methods in both high and low dimensional cases. For high-dimensional tuning, the advantage of ClassyTune over the other methods is more obvious. ClassyTune improves the performance of MySQL/TPC-C by more than \emph{six} times, while the GP-based BO and the search-based BestConfig can only improve by four times. Second, some systems can have only limited effective PerfConfs, e.g., PostgreSQL/TPC-C. The performance improvements are similar for both high and low dimensional tuning. Anyhow, ClassyTune still has a slight advantage over the other auto-tuning methods.

\subsection{Tuning Time}

\begin{figure}[!t]
\vspace{-9pt}
\centering
 \subfloat[ClassyTune outperforms other auto-tuning methods.]{
       \includegraphics[width=.23\textwidth,height=68pt]{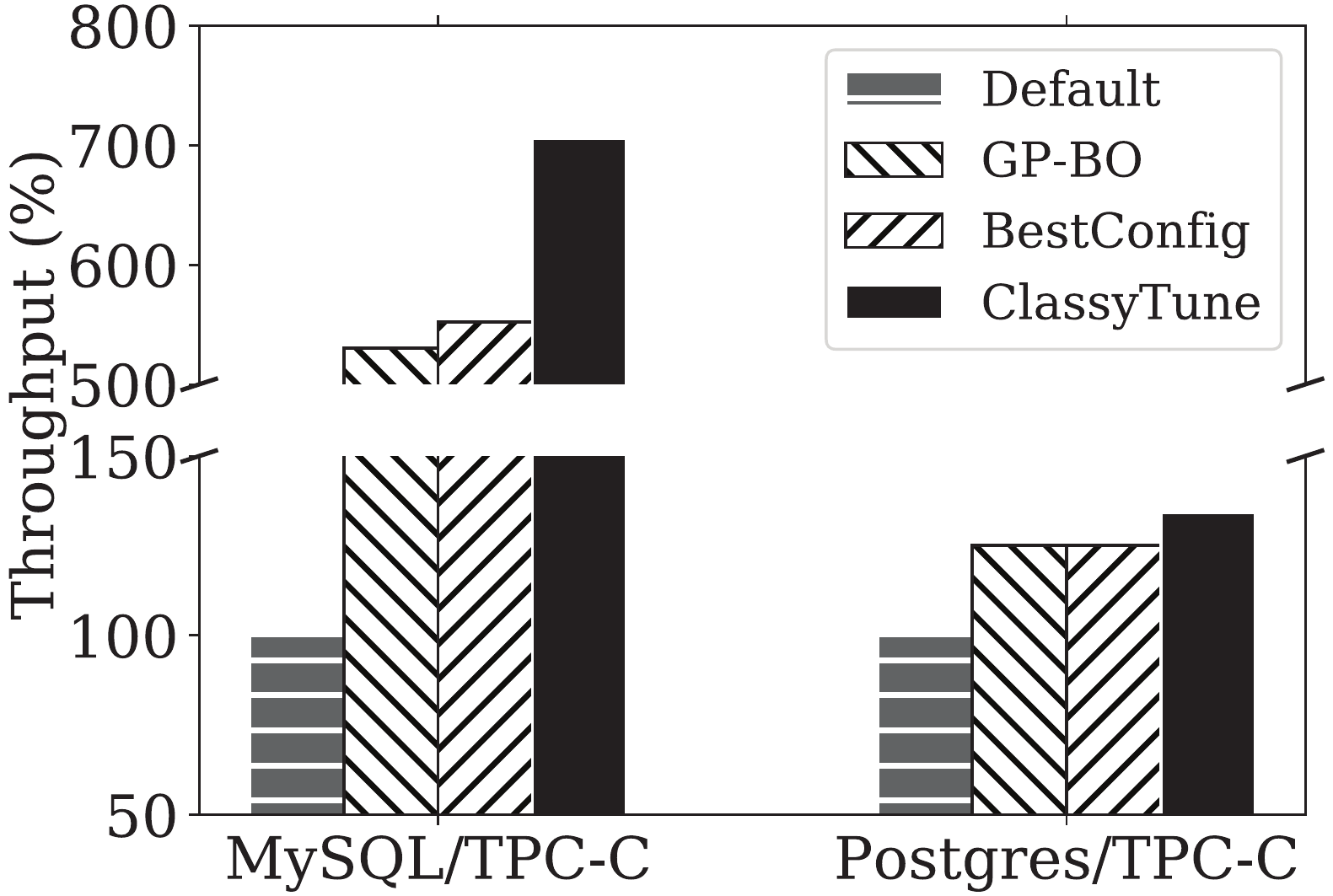}%
        \label{fig:hiditune:perf} 
    }\vspace{-3pt}
  \subfloat[Total tuning times for ClassyTune and GP-BO respectively.]{
        \includegraphics[width=.23\textwidth,height=68pt]{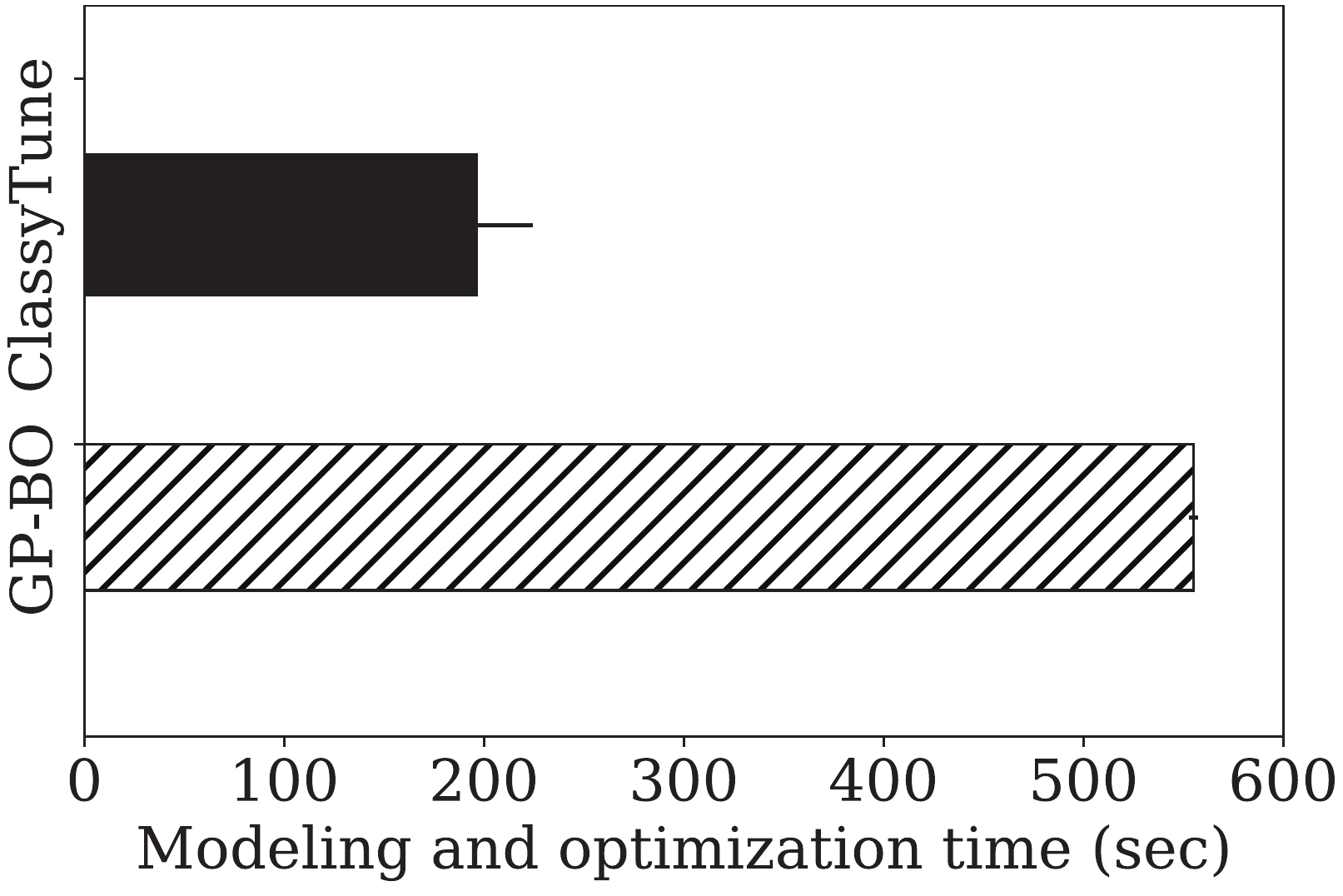}%
        \label{fig:hiditune:time} 
    }\vspace{-3pt}
     \caption{High-dimensional tuning results: tuning 30 PerfConfs for databases/TPC-C.}
     \label{fig:hiditune}
\end{figure}

We have mentioned that the GP-based GO method has high computation overhead. For the tuning results in Figure~\ref{fig:hiditune:perf}, we record the tuning time for both ClassyTune and GP-based BO. The tuning time includes the time for model training and model optimization. As GP-based BO is a stepwise method, the tuning time sums up all the computation time in all steps. We carry out the auto-tuning process of ClassyTune and GP-based BO for five times respectively. We report the average of the tuning results and the tuning times respectively. The results are plotted in Figure~\ref{fig:hiditune:time}.

ClassyTune involves a tuning time of no more than 200 seconds, while GP-based BO requires a tuning time of more than 550 seconds. Within a much shorter tuning time, ClassyTune finds a better PerfConf setting than the GP-based BO method. The GP-BO method has a heavy computation overhead because its tuning process involves the covariance matrix computation and this computation is carried out stepwise. Taking an integral approach to auto-tuning, ClassyTune trains a model once and then spends the rest of its time in searching the input space thoroughly based on the trained model. If necessary, ClassyTune can further reduce its tuning time by searching fewer points.\vspace{-6pt}
\subsection{Cloud Resource Reduction via Performance Tuning}
\label{sec:usecase}

ClassyTune can bring about the five benefits of automatic performance tuning~\cite{acts} like related works~\cite{bestconfig,ottertune,smartconf,smarthillclimbing}. Here, we present a real use case of UTuned's customers to show how ClassyTune enables cloud resource reduction via performance tuning. In this case, ClassyTune is used to tune a small online querying service deployed in the cloud. The application workload accesses the service by connecting to a stateless Web service cluster running a Spring Boot\footnote{\scriptsize https://spring.io/projects/spring-boot} application, which sends user queries to the backend. Before tuning, the service is deployed on a three-node cluster, supporting a throughput around 9000 composite operations per second. There is a resource planning question about \emph{whether all the three nodes are needed or reducing one node is possible, if the workload throughput must be guaranteed}.

To answer this resource planning question, we deploy the service on clusters of one to three nodes respectively. For each deployment, we test its performance under the default PerfConf setting. Then, we use ClassyTune to tune for the best performance. Under the tuned PerfConf setting, we test the service performance again. All the performance results are listed in Table~\ref{tbl:case}.
\begin{table}[!h]
    \vspace{-12pt}
	\caption{Service Throughputs: Default vs. Tuned}\vspace{-6pt}
    \label{tbl:case}%
    \centering
	\begin{tabular}{|c|c|c|}
		\hline
        \textbf{Node \#} & \textbf{Default} (err. rate)& \textbf{ClassyTune}  (err. rate)\\ \hline
		\textbf{1} & 3647.4 (17.6\%) & 4376.8 (8.9\%) \\ \hline
        \textbf{2} & 7972.4 (9.9\%) & \textbf{\normalsize 9341.3} (5.4\%) \\ \hline
        \textbf{3} &\textbf{\normalsize 9295.1} (9.7\%) & 11905.2 (2.2\%) \\ \hline
	\end{tabular}
\end{table}\vspace{-6pt}

For the target workload, a two-node cluster with a well-tuned PerfConf setting is the most cost effective. Without tuning, it would require one more node, i.e., 50\% more computing resources, to satisfy the application workload. From Table~\ref{tbl:case}, we can see that a one-node deployment, tuned or untuned, cannot support the application workload. While an untuned two-node deployment cannot meet the throughput requirement, it can perfectly support the workload after being tuned by ClassyTune. For three-node deployment, performance tuning enables it to support an even heavier application workload. In sum, We have actually reduced the cloud resource requirements (and \emph{costs}) of an online service by 33\% through performance tuning by ClassyTune.\vspace{-6pt}

\section{Conclusion}

This paper proposes a data-driven auto-tuning system ClassyTune, which can auto-tune the system performance by adjusting the PerfConfs within a limited number of tuning tests. ClassyTune exploits and models the comparison relations between PerfConfs by classification algorithms, instead of the typical performance-based model. Thanks to the comparison-based modeling, we can induce and generate more samples for training the classification model. Like other machine learning models, the classification model is not a hundred percent correct. If exploited naively, the imprecision of model could divert the performance tuning process such that no better PerfConf can be found. To guarantee a best PerfConf setting be found, we propose a clustering-based approach towards auto-tuning, exploiting the imprecise classification model.

Extensive experiments on seven systems commonly used in the cloud show that ClassyTune can outperform expert tuning and the state-of-the-art auto-tuning solutions, especially for high-dimensional inputs, while the computation overhead of ClassyTune is much lighter than that of the state-of-the-art GP-based BO method. An illustrative use case is presented to show how performance tuning by ClassyTune improves the system performance and enables the reduction of 33\% cloud computing resources for an online stateless service.



\ifCLASSOPTIONcaptionsoff
  \newpage
\fi


\balance 

\bibliographystyle{IEEEtran}
\bibliography{IEEEabrv,ref}

\end{document}